\documentclass[letterpaper,11pt]{article}
\pdfoutput=1

\usepackage{jheppub}

\usepackage{multirow}

\usepackage{subfig}
\usepackage{xspace}
\usepackage{xcolor}
\usepackage[countmax]{subfloat}

\definecolor{darkblue}{rgb}{0,0,0.5}

\definecolor{darkgreen}{rgb}{0,0.5,0}

\DeclareRobustCommand{\Sec}[1]{Sec.~\ref{#1}}
\DeclareRobustCommand{\Secs}[2]{Secs.~\ref{#1} and \ref{#2}}
\DeclareRobustCommand{\App}[1]{App.~\ref{#1}}

\DeclareRobustCommand{\Fig}[1]{Fig.~\ref{#1}}

\DeclareRobustCommand{\Eq}[1]{Eq.~(\ref{#1})}
\DeclareRobustCommand{\Eqs}[2]{Eqs.~(\ref{#1}) and (\ref{#2})}
\DeclareRobustCommand{\Ref}[1]{Ref.~\cite{#1}}
\DeclareRobustCommand{\Refs}[1]{Refs.~\cite{#1}}

\title{
New Insights on an Old Problem:\\Resummation of the $D$-parameter
}

\author{Andrew J. Larkoski}
\affiliation{Physics Department, Reed College, Portland, OR 97202, USA}
\emailAdd{larkoski@reed.edu}
\author{and Aja Procita}
\emailAdd{aja.procita@reed.edu}

\abstract{
The $D$-parameter is one of the oldest and most experimentally well-studied hadronic observables for $e^+e^-$ collisions.  Nevertheless, unlike other classic observables like the $C$-parameter or thrust, the $D$-parameter has never been resummed throughout its entire singular phase space.  Using insights and techniques motivated by modern multi-differential jet substructure calculations, we are able to predict the $D$-parameter distribution with no additional phase space cuts.  Our approach is to measure both the $C$- and $D$-parameters on hadronic final states in $e^+e^-$ collisions.  We can tune the value of the $C$-parameter with respect to the $D$-parameter to specify simple, physical configurations of final state particles in which to perform calculations.  There are three parametric regions that exist: $D \ll C^2\sim 1$, $D\ll C^2\ll 1$, and $D\sim C^2\ll1$, and we calculate the $D$-parameter in each region separately.  In the first two of these three regions, we present all-orders factorization theorems and explicitly demonstrate resummation to next-to-leading logarithmic accuracy.  The region in which $D\sim C^2\ll1$ corresponds to the dijet limit and where the $D$-parameter loses the property of additivity.  In this region we introduce a systematically-improvable procedure exploiting properties of conditional probabilities and resum to approximate next-to-leading logarithmic accuracy.  The contributions from these regions can be consistently combined, and the value of the $C$-parameter integrated over to produce the cross section for the $D$-parameter.  With these results, we match to leading fixed order as proof of principle and compare our resummed and matched prediction to data from LEP.
}

\begin{document}
\maketitle

\section{Introduction}

The legacy of the event shape measurements from experiments at the Large Electron-Positron Collider (LEP) on precision QCD calculations is profound.  Long after the experiments at LEP collected their last data, new frontiers of fixed-order and resummed perturbation theory have been established and used to study that data.  Programs for numerical evaluation of $e^+e^-\to$ three-jet scattering to next-to-next-to-leading fixed order (NNLO) now exist \cite{Gehrmann-DeRidder:2007nzq,DelDuca:2016csb,DelDuca:2016ily}, automated next-to-next-to-leading logarithmic order (NNLL) resummation for general observables is available \cite{Banfi:2014sua,Banfi:2016zlc}, and resummation for particular observables has even been extended to next-to-next-to-next-to-leading logarithmic order (N$^3$LL) \cite{Becher:2008cf,Abbate:2010xh,Chien:2010kc,Hoang:2014wka,Moult:2018jzp}.  These advances have enabled extractions of the strong coupling at the $Z$-pole, $\alpha_s(m_Z)$, with resummation at NNLL or N$^3$LL matched to NNLO \cite{Abbate:2010xh,Gehrmann:2012sc,Hoang:2015hka} that are now included in world averages provided by the Particle Data Group \cite{Patrignani:2016xqp}.

All the observables that have been calculated at this high precision, such as thrust \cite{Farhi:1977sg}, $C$-parameter \cite{Parisi:1978eg,Donoghue:1979vi,Ellis:1980wv}, and broadening \cite{Rakow:1981qn,Ellis:1986ig,Catani:1992jc}, have the property that they are first nonzero at tree-level for $e^+e^-\to q\bar q g$ (at ${\cal O}(\alpha_s)$).  This is vital, especially for resummation.  In the limit where resummation for thrust $\tau$ becomes important, $\tau \to 0$, all radiation in the event is constrained to be collinear with the hard final state quarks or are soft (low energy).  By momentum conservation, the hard quarks are back-to-back and the soft radiation is emitted off of the dipole formed from these quarks.  This physical configuration is simple and it is not necessary to include hard gluons sourcing jet structure for precision resummation.  Also, for these types of observables, including diagrams up through ${\cal O}(\alpha_s^3)$ corresponds to NNLO.  Indeed, while it can calculate in principle arbitrary observables, the program {\tt EERAD3} by default only calculates observables which are first nonzero for $e^+e^-\to q\bar q g$ \cite{Ridder:2014wza}.

In addition to these ${\cal O}(\alpha_s)$ observables, the experiments at LEP measured a number of observables that are sensitive to more exotic particle configurations.  In particular, because $e^+e^-$ scattering occurs in the center-of-mass frame, several observables that are sensitive to the aplanarity of the final hadronic system have been defined and measured.  These include observables like the thrust-minor \cite{Petersen:1987bq} and the $D$-parameter \cite{Parisi:1978eg,Donoghue:1979vi,Ellis:1980wv}.  Because these observables are only non-zero if the momentum of all final state particles is not coplanar, they are first non-zero at ${\cal O}(\alpha_s^2)$.  As such, fixed-order calculations of these observables cannot yet reach NNLO precision, and their resummation requires analysis of many different phase space regions that can contribute.  For example, a measurement of the $D$-parameter in the resummation regime $D\ll 1$ only constrains out-of-plane emissions.  The configuration of the particles in the plane are completely unconstrained, as long as they remain in the plane.  Because of this, theoretical studies of these observables had been restricted to fixed-order \cite{Nagy:1997yn,Campbell:1998nn} or resummation in the region in which there are three planar, well-separated jets, with soft or collinear emissions out of the plane \cite{Banfi:2000si,Banfi:2000ut,Banfi:2001sp,Banfi:2001pb}.  This can be accomplished by, for example, requiring that the three-jet resolution variable $y_3$ \cite{Catani:1991hj} is relatively large.  However, such a procedure does not produce an inclusive prediction for these observables because the measured value of $y_3$ restricts radiation.  Therefore, these predictions cannot be directly compared to the LEP measurements that had no such restriction.

In this paper, we present the first inclusive resummation calculations for these observables sensitive to aplanarity.  Our analysis will focus exclusively on the calculation of the $D$-parameter, though the procedure should be applicable to a wide range of similar observables.  This result is possible due to recent advances in resummation of multi-differential cross sections, which has seen extensive development in the field of jet substructure (see \Ref{Larkoski:2017jix} for a review).  A fundamental problem in the field of jet substructure is the construction of optimal observables for discrimination between jets of different origins; for example, discrimination of QCD jets initiated by light partons from jets from hadronic decays of highly-boosted $W$, $Z$, or $H$ bosons.  One difference between these jets is their prong structure: jets from decays of electroweak bosons will dominantly have a two-prong structure, while QCD jets will dominantly have only one prong.  There are many ways that a jet can manifest having two prongs, and so one must carefully identify the different phase space regions and then sum them up consistently.  This is the approach we take here to the resummation of the $D$-parameter: we will isolate the distinct phase space regions that can contribute to the $D$-parameter and then sum them up.

As mentioned earlier, to identify the distinct regions that contribute to the $D$-parameter, it is not sufficient to just measure the $D$-parameter.  This is in contrast to thrust, for example, where the value of thrust regulates all singular configurations.  So, with insight from jet substructure calculations, we will measure the $D$-parameter and another observable on the final state which will control the planar configuration of particles.  As this auxiliary observable, we choose to measure the $C$-parameter.  This choice is due to the relation of the $C$- and $D$-parameters as defined by the spherocity tensor, but one could choose another observable to accomplish the same task.  For the resummation regime when $D\ll 1$, we will demonstrate that there are three regions defined by the relationship between the values of $C$ and $D$:
\begin{enumerate}

\item Region 1 (Trijets): $D \ll C^2 \sim 1$.  This is the region where the $C$-parameter is well described at tree-level fixed-order because the final state of $e^+e^-\to q\bar q g$ is planar, in the center-of-mass frame.  This region was studied and resummed in \Ref{Banfi:2001pb}, though using the three-jet resolution variable $y_3$ to define the planar configuration.

\item Region 2 (Hierarchical Trijets): $D \ll C^2 \ll1$.  The radiation that sets the $D$-parameter in this region is still parametrically separated from the radiation that sets the $C$-parameter.  However, now, the value of the $C$-parameter must be resummed as well, as there will be large logarithms of $C$ in addition to $D$ in this region.

\item Region 3 (Dijets): $D\sim C^2 \ll 1$.  In this region, logarithms of $C$ must be resummed, but there will be no additional logarithms of the $D$-parameter.  The physical configuration of this region corresponds to the dijet limit, for which the $D$-parameter is no longer an additive observable.

\end{enumerate}
There are no regions with more than three resolved jets that contribute at leading power in the limit that $D\ll1$.  With more than three jets in the final state, they are no longer constrained to lie in a plane in the center-of-mass frame.  The existence of more jets in the final state does not produce any large logarithms, and enforcing $D\ll 1$ requires that all the jets lie within an angle of order the value of $D$ of the plane.

To accomplish resummation of the $C$- and $D$-parameters (as appropriate) in the first two of these three regions, we construct effective theories in each region with the formalism of soft-collinear effective theory (SCET) \cite{Bauer:2000yr,Bauer:2001ct,Bauer:2001yt,Bauer:2002nz}.  The general framework for resummation with multi-jet hierarchies was presented in \Ref{Pietrulewicz:2016nwo} with an effective theory called SCET$_+$.  We factorize the cross section differential in $C$ and $D$ into hard, collinear, and soft contributions that are each defined at natural scales.  Resummation is then accomplished by renormalization group evolution between the hard, collinear, and soft scales.  While our factorization will be valid to any logarithmic accuracy, for concreteness we will present results at next-to-leading logarithmic accuracy (NLL).  By ``NLL'' we mean that we are able to successfully resum logarithms of the form $(\alpha_s\log D)^n$ in the exponent of the cumulative distribution of the $D$-parameter.  In the language of the factorization theorem, we calculate and resum with one-loop anomalous dimensions.

While we calculate all necessary pieces in regions 1 and 2 for complete resummation to NLL accuracy, for the plots we present in this paper, we will combine them with a simple prescription that resums all but one source of logarithms relevant for NLL.  The factorized cross section valid in regions 1 and 2 where $D\ll C^2$ can be expressed in the compact form of
\begin{align}
\frac{1}{\sigma}\frac{d^3\sigma^{D\ll C^2}}{dD\, dx_1\, dx_2} = H(Q^2)H_{(2\to 3)}(x_1,x_2)J_1(x_1,D)J_2(x_2,D)J_3(x_3,D)S_{123}(x_1,x_2,D)\,.
\end{align}
Here, $H(Q^2)$ is the hard function for production of dijets in $e^+e^-$ collisions, $J_i(x_i,D)$ is the jet function describing collinear radiation off of jet $i$ in the final state, and $S_{123}(x_1,x_2,D)$ is the soft function that describes low-energy gluon emission off of the dipoles formed from pairs of final state jets.  $H_{(2\to 3)}(x_1,x_2)$ is the hard function describing the formation of the third jet; its presence and resummation in this factorization theorem is what enables a consistent combination of regions 1 and 2.  The $x_i$ are the three-body phase space variables which in the center-of-mass frame are just energy fractions.  The $C$-parameter in these phase space regions is just a function of the $x_i$.  The jet and soft functions are implicitly convolved with one another.  This expression is valid for all values of $x_1$ and $x_2$ such that they enforce $D\ll C^2$ and only misses contributions from logarithms of the $D$-parameter from collinear-soft radiation \cite{Bauer:2011uc}.  

Because the $D$-parameter is non-additive in the third region, resummation via factorization is confounded and so we take a different approach.  As there is no hierarchy between $C$ and $D$, only large logarithms of $C$ need to be resummed which suggests a way to split the double differential cross section in this region.  Recognizing the double differential cross section as a joint probability distribution, we can express it with the definition of a conditional probability:
\begin{equation}
\frac{d^2\sigma^{D\sim C^2\ll 1}}{dC\, dD}=\frac{d\sigma}{dC}\frac{d\sigma(C)}{dD}\,.
\end{equation}
Here, $d\sigma(C)/dD$ is the differential cross section of $D$ conditioned on the value of $C$.  As written this is an identity: the equality is exact if everything is calculated to the same accuracy.  However, in this form, this enables the cross section of $C$ and the conditional cross section of $D$ to be calculated at different accuracy for an approximate result.  We use results from the literature for the NLL resummed cross section of the $C$-parameter, and calculate the conditional cross section to lowest non-trivial order, ${\cal O}(\alpha_s^2)$.  This produces an approximate NLL accurate double differential cross section in this region which, at any rate, is formally accurate to fixed-order and systematically improvable.

With the resummed cross section calculated in each of these regions, we then just need to match them to construct a cross section that is differential in $D\ll 1$, for any value of $C$.  Region 3 can then be included by a simple additive matching, appropriately subtracting the overlap with regions 1 and 2.  Once this super-cross section has been constructed, then the cross section for the $D$-parameter exclusively is found by integrating over the value of $C$:
\begin{equation}
\frac{d\sigma}{dD} = \int_{\left(
\frac{4}{3}D
\right)^{1/2}}^{3/4} dC\, \frac{d^2\sigma}{dC\, dD}\,.
\end{equation}
The upper bound of the integral over $C$ is $3/4$, as that is the maximum value for the $C$-parameter when $D\ll 1$.  The lower bound of this integral is $\left(\frac{4}{3}D\right)^{1/2}$ as this is the minimum value that the $C$-parameter can take given a value of $D$.  We will show both of these limits in the next section.  This procedure then produces a resulting resummed cross section for the $D$-parameter that is completely inclusive of any other restrictions on the event.

The outline of this paper is as follows.  In \Sec{sec:obs}, we define the $C$- and $D$-parameter observables, and present useful, but perhaps non-standard, expressions for them.  In \Secs{sec:reg1}{sec:reg2}, we present the factorized expressions for the cross section in regions 1 and 2.  By appropriately defining scales in the functions of these factorization theorems, we can combine these regions, which is done in \Sec{sec:comb}.  The region 3 calculation is presented in \Sec{sec:reg3}, which has a unique form because the $D$-parameter is not additive in this region.  Our procedure for consistently combining the cross sections in the three regions is described in \Sec{sec:combine}.  \Sec{sec:np} describes the scaling of non-perturbative corrections to the $D$-parameter and the regions for which they are most important.  The factorization theorems and completeness of the three regions are validated in \Sec{sec:event2} by numerical comparison of our predictions to the fixed-order code {\tt EVENT2} \cite{Catani:1996vz} at leading-order, ${\cal O}(\alpha_s^2)$. The matching of fixed-order to our resummed cross section for $D$ is presented in \Sec{sec:lep}, along with a comparison to data from the experiments at LEP.  We conclude in \Sec{sec:concs}.  Appendices contain explicit calculations and other technical details of the factorization theorems.

\section{Observable Definitions}\label{sec:obs}

The $C$- and $D$-parameters are defined from the spherocity tensor (assuming all massless particles) \cite{Parisi:1978eg,Donoghue:1979vi,Ellis:1980wv}:
\begin{equation}
\Theta_{\alpha\beta} = \frac{1}{Q}\sum_i \frac{p_{i\alpha}p_{i\beta}}{E_i}\,.
\end{equation}
The sum runs over all particles in the final state of $e^+e^-$ collisions with $Q$ the center-of-mass collision energy, $E_i$ the energy of particle $i$, and $p_{i\alpha}$ is the $\alpha$ component of particle $i$'s three-momentum.  The $C$- and $D$-parameters are defined by the eigenvalues of the spherocity tensor $\lambda_i$, with the ordering $\lambda_1\geq \lambda_2\geq\lambda_3$ and $\text{tr}\, \Theta = \lambda_1+\lambda_2+\lambda_3=1$.  The $C$-parameter is
\begin{equation}\label{eq:cdef}
C=3(\lambda_1\lambda_2+\lambda_1\lambda_3+\lambda_2\lambda_3) = \frac{3}{Q^2}\sum_{i< j} E_iE_j\sin^2\theta_{ij}\,.
\end{equation}
The $D$-parameter is
\begin{equation}
D = 27\lambda_1\lambda_2\lambda_3\,.
\end{equation}
The normalization factors of 3 and 27, respectively, are such that the $C$- and $D$-parameters range from 0 to 1.  The $D$-parameter can also be written in terms of energies and relative angles as
\begin{align}\label{eq:ddef}
D&=\frac{27}{Q^3}\sum_{i<j<k} \frac{|{\bf p}_i\cdot ({\bf p}_j\times {\bf p}_k)|^2}{E_iE_jE_k}\\
&=\frac{27}{Q^3}\sum_{i<j<k} E_iE_jE_k(1+2\cos\theta_{ij}\cos\theta_{jk}\cos\theta_{ik}-\cos^2\theta_{ij}-\cos^2\theta_{jk}-\cos^2\theta_{ik})\,.\nonumber
\end{align}
The expression for $D$ on the second line of this equation will be especially useful, where $\theta_{ij}$ is the angle between particles $i$ and $j$.  Note that these identities hold when considering exclusively massless particles; we will assume this throughout this paper.

The possible hierarchical relationships between the eigenvalues of the spherocity tensor can be exploited to identify the different regions that we need to consider for resummation of the $D$-parameter.  First, if the eigenvalues are all of the same order:
\begin{equation}
1\sim\lambda_1 \gtrsim \lambda_2\gtrsim \lambda_3\,,
\end{equation}
then both $C$ and $D$ are of order-1, and are well-described at fixed-order.  Therefore, there is no resummation necessary with this scaling.

If there is one eigenvalue that is parametrically small than the others:
\begin{equation}
1\sim \lambda_1\gtrsim \lambda_2\gg \lambda_3\,,
\end{equation}
this corresponds to the physical configuration when the final state is nearly planar.  In this limit, the $C$-parameter reduces to 
\begin{equation}
C= 3\lambda_1 \lambda_2\sim 1\,,
\end{equation}
and has an order-1 value and is described at fixed-order.  The $D$-parameter, on the other hand, is parametrically smaller:
\begin{equation}
D = 27\lambda_1 \lambda_2\lambda_3 = 3\frac{\lambda_3}{\lambda_1\lambda_2}(3\lambda_1\lambda_2)^2= 3\frac{\lambda_3}{\lambda_1\lambda_2}C^2 \ll C^2\sim 1\,.
\end{equation}
Therefore $D\ll C^2\sim 1$ which corresponds to region 1 as defined above.  The resummation of the $D$-parameter in this phase space region was done to NLL accuracy in \Ref{Banfi:2001pb}.  Instead of using the value of the $C$-parameter, \Ref{Banfi:2001pb} used the three-jet resolution variable $y_3$ \cite{Catani:1991hj} to constrain the system to have three jets.  Note that the maximum value of the $C$-parameter in this region of phase space occurs when $\lambda_1=\lambda_2=1/2$, and so $C \leq 3/4$.

If the three eigenvalues are all parametrically separated:
\begin{equation}
1\sim \lambda_1\gg \lambda_2\gg \lambda_3\,,
\end{equation}
this corresponds to the physical configuration when the final state is nearly planar and approaching the dijet limit.  The $C$-parameter in this limit reduces to
\begin{equation}
C= 3\lambda_2\ll 1\,,
\end{equation}
as $\lambda_1\to 1$ in this limit.  The $D$-parameter is related to it as:
\begin{equation}
D = 27 \lambda_2\lambda_3 = 3\frac{\lambda_3}{\lambda_2}(3\lambda_2)^2= 3\frac{\lambda_3}{\lambda_2}C^2 \ll C^2\ll 1\,.
\end{equation}
Therefore, $D \ll C^2 \ll 1$, which corresponds to region 2 as defined above.  Restricting $D\ll C^2 \ll 1$ means that there are three jets in the final state, but one of them is soft, or collinear to another jet.  This configuration can be resummed using the factorization theorems of \Ref{Bauer:2011uc} (for the collinear jets) or \Ref{Larkoski:2015zka} (for the soft jet).  \Ref{Larkoski:2015kga} described how to combine them consistently for the particular application of the $D_2$ observable \cite{Larkoski:2013eya,Larkoski:2014gra}.  Note that because there are two hierarchies, one must resum both $C$ and $D/C^2$, which is accomplished in the factorization theorems referenced.  Also, in this limit the $D$-parameter is additive, so its resummation is ``simple.''

There is one final hierarchical relationship between the eigenvalues:
\begin{equation}
1\sim \lambda_1\gg \lambda_2\sim \lambda_3\,.
\end{equation}
In this limit, the $C$-parameter becomes
\begin{align}
C=3(\lambda_2+\lambda_3)\,,
\end{align}
while the $D$-parameter is
\begin{equation}
D = 27 \lambda_2\lambda_3\,.
\end{equation}
That is, $D\sim C^2 \ll 1$ which corresponds to the dijet limit, region 3.  Because of this relationship, the $D$-parameter is no longer additive, but there are no large logarithms of $D/C^2\sim 1$.  Therefore, in this phase space region, we only need to resum logarithms of $C\ll1$, which is still an additive observable.   Resummation of this form is like the ``soft haze'' region described in \Ref{Larkoski:2015kga}.

In this final region, it is useful to determine the precise relationship between the $C$- and $D$-parameters.  Note that the $D$-parameter can be arbitrarily smaller than the $C$-parameter as $\lambda_3\to 0$.  However, the maximum value that the $D$-parameter can take occurs when $\lambda_2=\lambda_3$.  In this limit, we have the relationship
\begin{equation}\label{eq:cdrel}
D\leq \frac{3}{4}C^2\,.
\end{equation}
This upper limit will be important to remember when constructing the contributions to the $D$-parameter in this region.

To summarize: in region 1, we only resum $D$ and calculate $C$ at fixed-order; in region 2, we need to resum both $C$ and $D$; and in region 3, we only need to resum $C$ and calculate $D/C^2$ at fixed-order.  We will now discuss in detail the form of the factorized cross section in each region.

\section{Region 1 (Trijets): $D \ll C^2\sim 1$}\label{sec:reg1}

We begin by constructing the cross section in region 1, when $D \ll C^2\sim 1$.  Because $C\sim 1$ in this region, the $C$-parameter is well-described at fixed-order by the production of three well-separated, energetic jets in the final state.  Additionally, because we assume that $D \ll 1$, the radiation in the final state that is out of the plane defined by those three jets is constrained to be at small angle with respect to the plane, or have low energy.  Those emissions that are at small angle to the plane could in principle be at any angle with respect to the energetic jets that set the value of the $C$-parameter.  However, the emission probability for this small-angle radiation is only enhanced by a large logarithm if it is additionally collinear to one of the energetic jets.  Therefore, to leading power, the emissions that set the value of the $D$-parameter are soft (low energy) or collinear to one of the three energetic jets that set the value of the $C$-parameter.

The cross section in this region of phase space factorizes just assuming hard-collinear-soft factorization of QCD.  The form of the factorized cross section in this region is
\begin{equation}\label{eq:factreg1}
\frac{1}{\sigma}\frac{d^3\sigma}{dD\, dx_1\, dx_2} = H_{3-\text{jet}}(x_1,x_2)J_1(x_1,D)\otimes J_2(x_2,D)\otimes J_3(x_3,D)\otimes S_{123}(x_1,x_2,D)\,.
\end{equation}
To write this cross section, we have used the three-body phase space variables $x_1$, $x_2$, and $x_3$ where
\begin{equation}
x_i=\frac{2E_i}{Q}\,.
\end{equation}
$E_i$ is the energy of the $i$th jet in the event, which can be operationally defined through an exclusive jet algorithm that finds three final-state jets.  However, this multi-differential cross section is not directly measurable in experiment because it requires identification of the quark and gluon jets that compose the final state.  Because $D\ll 1$, the masses of these jets are very small compared to their energies and so, to leading power, their phase space is completely defined by these energy fractions.  $Q$ is the total center-of-mass energy and $x_1+x_2+x_3=2$.  Because of conservation of energy, there are only two independent phase space variables, which we take to be $x_1$ and $x_2$, the energy fractions of the quark and anti-quark in the final state.  The tree-level expression for the 3-jet hard function $H_{3-\text{jet}}(x_1,x_2)$ is simply the leading-order matrix element for $e^+e^-\to q\bar q g$ divided by the cross section for $e^+e^-\to q\bar q$:
\begin{equation}
H^{(0)}_{3-\text{jet}}(x_1,x_2) = \frac{\alpha_s}{2\pi}C_F\frac{x_1^2+x_2^2}{(1-x_1)(1-x_2)}\,.
\end{equation}
The $C$-parameter as measured on this final state has the value
\begin{equation}
C=6\frac{(1-x_1)(1-x_2)(1-x_3)}{x_1 x_2 x_3}=6\frac{(1-x_1)(1-x_2)(x_1+x_2-1)}{x_1 x_2 (2-x_1-x_2)}\,.
\end{equation}
We refer the reader to \Refs{Ellis:2010rwa,Pietrulewicz:2016nwo} for matrix element definitions of hard functions for multi-jet production and soft functions with multiple Wilson lines.

The other functions in the factorization theorem describe emission of collinear radiation or soft radiation that contributes to the measured value of $D$.  In this region of phase space, the $D$-parameter can be expressed in terms of the out-of-plane momentum of a final state particle as \cite{Banfi:2001pb}
\begin{equation}\label{eq:dparamdefreg1}
D = 9 C \sum_{j}\frac{p_{j,\text{out}}^2}{E_j Q}\,,
\end{equation}
where the sum over $j$ runs over all final state particles.  Note that this is directly dependent on the value of $C$.  Importantly, this demonstrates that the $D$-parameter is an additive observable in this region; this will enable resummation to be accomplished in a relatively simple way.  At the level of the factorized cross section, additivity is imposed by convolution (denoted by $\otimes$) of the various functions that depend on the value of $D$.

The jet function $J_i(x_i,D)$ describes the production of radiation that is collinear to energetic jet $i$ and contributes to the value of the $D$-parameter.  Using the expression for the $D$-parameter as in \Eq{eq:dparamdefreg1}, the contribution to $D$ from collinear radiation off of this jet is
\begin{equation}\label{eq:dcollreg1}
D_i = \frac{9}{2}Cx_i \sum_{j\in \text{ jet }i}z_j \theta_j^2\sin^2\phi_j\,.
\end{equation}
Here, the sum runs over those collinear particles in jet $i$, $z_j$ is the energy fraction with respect to the jet energy, $\theta_j$ is the angle of the particle from the jet axis, and $\phi_j$ is the azimuthal angle about the in-plane jet axis.  This angle is 0 or $\pi$ if the particle is in the plane.  For a jet with two particles (sufficient for resummation to NLL accuracy), this simplifies to
\begin{equation}
D_i = \frac{9}{2} C x_iz(1-z)\theta^2\sin^2\phi\,,
\end{equation}
where now $z$ and $1-z$ are the energy fractions of the particles in the jet, and $\theta$ is their relative angle.

The soft function $S_{123}(x_1,x_2,D)$ describes low-energy radiation emitted off of the dipoles formed from any pair of the three final state jets.  It depends only on the relative angles between the jets, and therefore by momentum conservation on the energy fractions $x_i$.  The contribution to the $D$-parameter from soft emissions can be expressed as:
\begin{equation}\label{eq:dsoftreg1}
D = 9 C\sum_{j\text{ soft}} \frac{E_j}{Q}\sin^2\theta_{j} \sin^2\phi_j = 9C\sum_{j\text{ soft}}\frac{k_{\perp,j}}{Q}\frac{\sin^2\phi_j}{\cosh\eta_j}\,.
\end{equation}
In the first equality, we have written it in terms of the energy $E_j$ of the soft emission $j$ and its angle $\theta_j$ from one of the hard jets and its angle $\phi$ above the plane of the three hard jets.  In the second equality, we have changed variables to express $D$ in terms of the momentum $k_{\perp,j}$ of soft emission $j$ out of the plane, angle $\phi_j$ out of the plane, and rapidity $\eta_j$ with respect to one of the hard jet directions.  

\begin{figure}[t]
\begin{center}
\includegraphics[width=.45\textwidth]{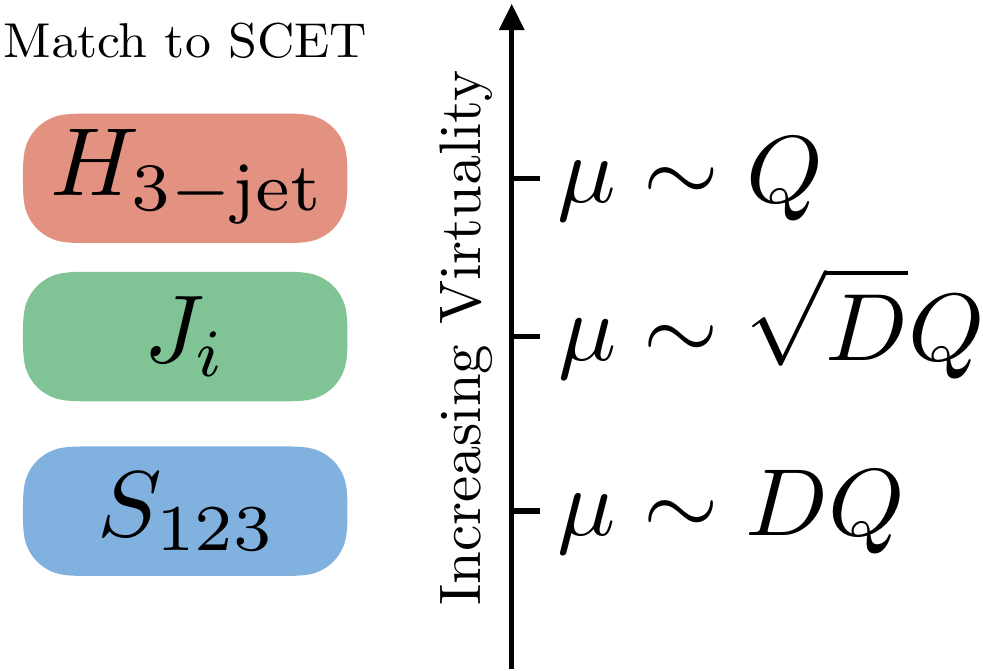}
\end{center}
\caption{
Illustration of the natural scales of the functions appearing in the factorization theorem of the trijets phase space region.
}
\label{fig:reg1scales} 
\end{figure}

To resum this factorized cross section to NLL accuracy in the $D$-parameter, we must calculate the one-loop anomalous dimensions of the functions in the factorization theorem.  The calculation of the anomalous dimensions in this phase space region are presented in \App{app:reg1}.   The relative scales of the functions appearing in the factorization theorems are displayed in \Fig{fig:reg1scales}.  Because this factorization theorem only required hard-soft-collinear factorization, this scales plot doesn't encode much information.  However, when considering the other relevant phase space regions it will be important, and so is a benchmark for additional scales that appear in those cases.  With this resummed and factorized form of the cross section, we can then calculate the cross section differential in the $C$- and $D$-parameters by marginalizing over $x_1$ and $x_2$:
\begin{equation}
\frac{d^2\sigma}{dC\, dD} = \int_0^1 dx_1 \int_0^1 dx_2\, \Theta(x_1+x_2-1)\frac{d^3\sigma}{dD\, dx_1\, dx_2}\delta\left(
C-6\frac{(1-x_1)(1-x_2)(x_1+x_2-1)}{x_1 x_2 (2-x_1-x_2)}
\right)\,.
\end{equation}
While the $C$-parameter $\delta$-function integral can be done in principle, in practice, this integral is easiest to do by Monte Carlo.

\section{Region 2 (Hierarchal Trijets): $D \ll C^2\ll 1$}\label{sec:reg2}

The hierarchal jets region, region 2, corresponds to the relationship $D \ll C^2 \ll 1$.  Because now both $C$ and $D$ are parametrically less than 1, we must resum both observables.  This is accomplished through a re-factorization of the cross section presented in the previous section.  There are two possible configurations that must be considered in this region.  At leading power, this corresponds to either the gluon jet becoming collinear with one of the quark jets, or the gluon jet becoming soft as compared the center-of-mass energy.  Because we still require $D \ll C^2$, this means that the final state is still parametrically more planar than it is dijet-like.  This will enable a direct connection with the factorization theorem of region 1, which we will discuss in the next section.  We will address these two configurations separately.  The form of the factorization theorems presented in this section and the method for their consistent combination was first discussed in \Ref{Larkoski:2015kga} and the general jet hierarchy configuration was analyzed in \Ref{Pietrulewicz:2016nwo}.

\subsection{Collinear Gluon Jet}

For concreteness, consider the configuration where the gluon becomes collinear with the anti-quark.  In terms of the three-body phase space variables, this corresponds to the limit $x_1\to 1$.  In this limit, the expression for the $C$-parameter simplifies greatly:
\begin{equation}\label{eq:collx1}
C \to 6(1-x_1)\,.
\end{equation}
The expression for the $D$-parameter is modified appropriately in this limit, and still has contributions from emissions collinear to and at low energy with respect to the hard(er) jets in the final state.  The factorized cross section in this region of phase space was first studied in \Ref{Bauer:2011uc}.  Here, we present the derivation of the factorization theorem in this phase space region as a re-factorization of the factorization theorem of \Eq{eq:factreg1}. 

Starting from that factorized cross section,
\begin{equation}
\frac{1}{\sigma}\frac{d^3\sigma}{dD\, dx_1\, dx_2} = H_{3-\text{jet}}(x_1,x_2)J_1(x_1,D)\otimes J_2(x_2,D)\otimes J_3(x_3,D)\otimes S_{123}(x_1,x_2,D)\,,
\end{equation}
we then take the limit that $x_1\to 1$.  In this limit, the jet functions $J_i$ are not re-factorized; they just take their leading-power value with $x_1\to 1$.  The hard function $H_{3-\text{jet}}$, however, must be re-factorized because there is a new hierarchy introduced by $x_1\to 1$.  In this limit, the hard function factorizes as
\begin{equation}
H_{3-\text{jet}}(x_1,x_2) \to H_{2-\text{jet}}(Q^2)H_{1\to 2}(x_1,z)\,.
\end{equation}
Here, $H_\text{2-jet}$ is the hard function for dijet production in $e^+e^-$ collisions with center-of-mass energy $Q$ which is just 1 at tree-level.  $H_{1\to 2}$ is the quark collinear splitting function, which is the $x_1\to 1$ limit of the tree-level $e^+e^-\to q\bar q g$ cross section:
\begin{equation}
H_{1\to 2}^{(0)}(x_1,z) = \lim_{x_1\to 1} \left.\frac{\alpha_s}{2\pi}C_F\frac{x_1^2+x_2^2}{(1-x_1)(1-x_2)}\right|_{1-x_2\to z} = \frac{\alpha_s}{2\pi}C_F\frac{1}{1-x_1}\frac{1+(1-z)^2}{z}\,.
\end{equation}
We have denoted the energy fraction of the gluon in the splitting as $z$ where
\begin{equation}
z = \frac{E_g}{E_g+E_{\bar q}} = \frac{x_3}{x_2+x_3} = 1-x_2\,.
\end{equation}

The soft function also re-factorizes because the angle between two of the jets that source soft radiation is becoming small.  Honest wide-angle soft radiation can only resolve the total $q\bar q$ dipole, while soft radiation boosted along the collinear anti-quark and gluon direction can resolve their splitting.  Therefore, the soft function factorizes into these two components:
\begin{equation}
S_{123}(x_1,x_2,D)\xrightarrow[x_1\to 1]{} S_{12}(C,D)\otimes C_s(C,D,x_2)\,.
\end{equation}
Here, $S_{12}(C,D)$ is the soft function for radiation off of the $q\bar q$ dipole and $C_s(C,D,x_2)$ is the collinear-soft function that describes soft radiation that can resolve the anti-quark--gluon collinear splitting.

\begin{figure}[t]
\begin{center}
\includegraphics[width=.65\textwidth]{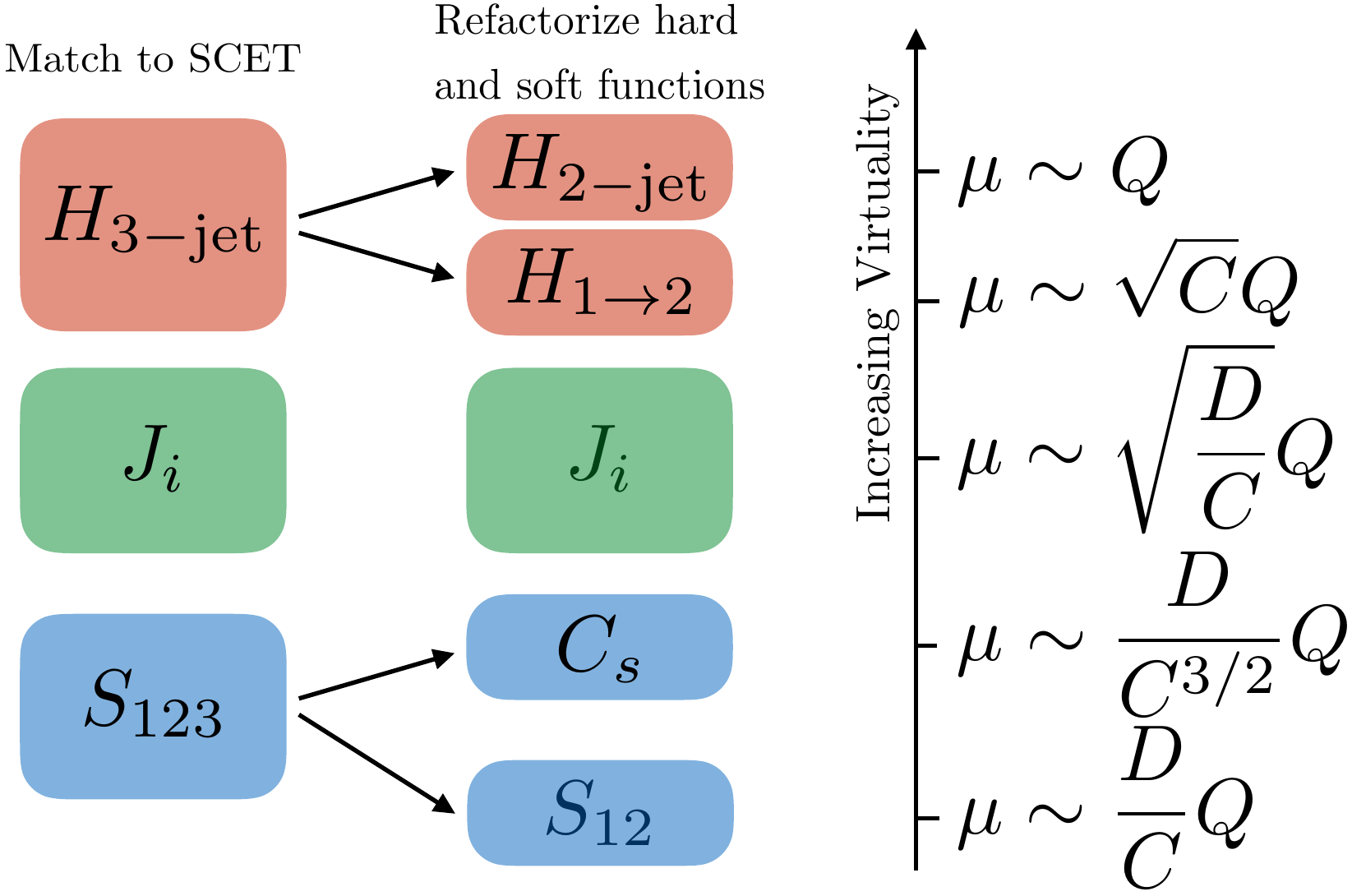}
\end{center}
\caption{
Illustration of the re-factorization and natural scales of the functions appearing in the factorization theorem of the collinear hierarchal trijets phase space region.
}
\label{fig:reg2collscales} 
\end{figure}

This cross section is more naturally expressed in terms of the energy fraction $z$ and the splitting angle $\theta$ of the gluon off of the anti-quark.  In terms of the three-body phase space variables, note that the invariant mass of the anti-quark--gluon system is
\begin{equation}
2p_g\cdot p_{\bar q} = 2E_g E_{\bar q} (1-\cos\theta) = \frac{Q^2}{2}x_2 x_3(1-\cos\theta) = Q^2(1-x_1)\,.
\end{equation}
In the limit where $x_1\to 1$, the splitting angle $\theta$ is therefore
\begin{equation}
\theta^2 = \frac{4(1-x_1)}{z(1-z)}\,.
\end{equation}
This enables an equivalent representation of the $C$-parameter in this limit:
\begin{equation}\label{eq:ccoll}
C = \frac{3}{2}z(1-z)\theta^2\,.
\end{equation}
The fully factorized cross section is therefore
\begin{equation}
\frac{1}{\sigma}\frac{d^3\sigma}{dD\, dC\, dz} = H_{2-\text{jet}}(Q^2)H_{1\to 2}(z,C)J_1(C,D)\otimes J_2(1-z,D)\otimes J_3(z,D)\otimes S_{12}(C,D)\otimes C_s(C,D,z)\,.
\end{equation}
In this expression, we have used \Eqs{eq:collx1}{eq:ccoll} to express the cross section exclusively in terms of the gluon energy fraction $z$, $C$, and $D$.  In \Fig{fig:reg2collscales} we show the re-factorization and hierarchy of scales in this factorization theorem.  The anomalous dimensions and tree-level expressions of the hard and soft functions are given in \App{app:collreg2anom}.  The anomalous dimensions of the jet functions are found from taking limits of the anomalous dimensions presented in \App{app:anomsumm}.  There is an essentially identical region when $x_2\to 1$ which is just found with the replacement $x_1\leftrightarrow x_2$ in this factorization theorem.

\subsection{Soft Gluon Jet}

The phase space region where the gluon becomes soft corresponds to the limit in which the three-body phase space coordinates become:
\begin{equation}
x_1\,,x_2\to 1\,.
\end{equation}
Note that this corresponds to the limit in which the gluon energy fraction $x_3\to 0$.  In this limit, the $C$-parameter becomes
\begin{equation}
C\to 6 \frac{(1-x_1)(1-x_2)}{2-x_1-x_2}\,.
\end{equation}
Just like in the collinear jet region, the expressions for the $D$-parameter are appropriately modified from its value in region 1, but we won't show the explicit expressions here.  The factorization of this soft jet region of phase space was first studied in \Ref{Larkoski:2015zka}.  As with the collinear jets, we will present the refactorization necessary to derive the cross section in this region.

In this soft gluon region, the jet functions $J_i$ and soft function $S_{123}$ are just modified according to the leading expression in the $x_1,x_2\to 1$ limit.  This is true because of the assumed hierarchy of $D\ll C^2\ll 1$, and so the final state is more co-planar than it is dijet-like.  The hard function $H_{3-\text{jet}}$ factorizes into the $e^+e^-\to $ dijets function $H_{2-\text{jet}}$ and the function that creates a soft gluon at an arbitrary angle from the dijets, $H_s$:
\begin{equation}
H_{3-\text{jet}}(x_1,x_2) \to H_{2-\text{jet}}(Q^2)H_s(z,C)\,.
\end{equation}
Here, $z$ is the energy fraction of the soft gluon with respect to the total collision energy,
\begin{equation}
z = \frac{E_g}{Q} = \frac{x_3}{2}\,.
\end{equation}
The tree-level expression for $H_s$ is found by taking the appropriate limit of the $e^+e^-\to q\bar g$ matrix element
\begin{align}
H_s^{(0)}(z,C) &= \lim_{x_1,x_2\to 1} \left.\frac{\alpha_s}{2\pi}C_F\frac{x_1^2+x_2^2}{(1-x_1)(1-x_2)}\right|_{1-x_3/2\to z} = \frac{\alpha_s}{\pi}C_F\frac{1}{(1-x_1)(1-x_2)}\\
&=\frac{\alpha_s}{\pi}C_F\frac{2}{z\sin^2\theta}\,.
\nonumber
\end{align}
In the second line, we have expressed the cross section in terms of the energy fraction $z$ of the gluon and the angle from the quark--anti-quark pair, $\theta$.

\begin{figure}[t]
\begin{center}
\includegraphics[width=.65\textwidth]{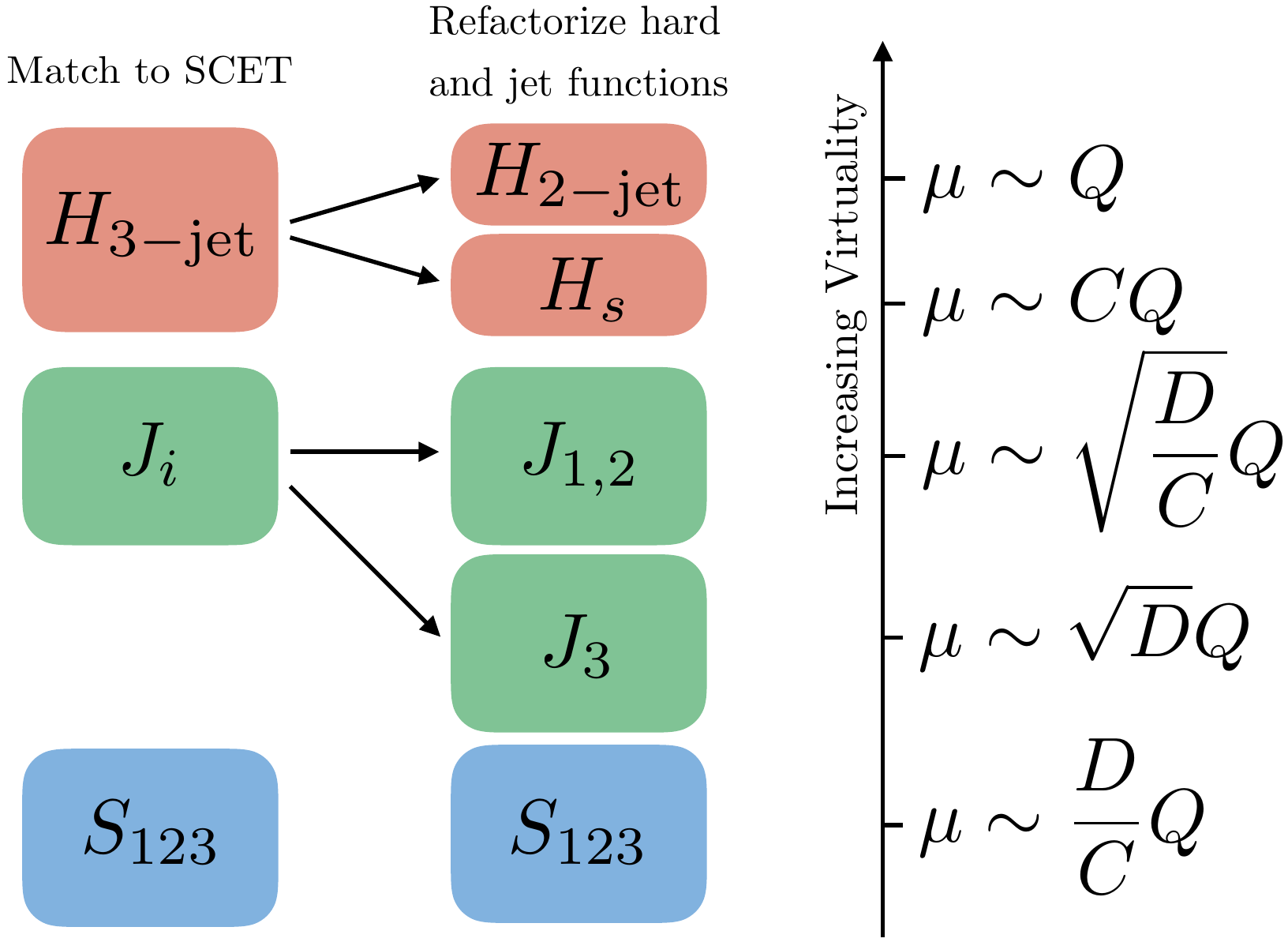}
\end{center}
\caption{
Illustration of the re-factorization and natural scales of the functions appearing in the factorization theorem of the soft hierarchal trijets phase space region.
}
\label{fig:reg2softscales} 
\end{figure}

The cross section in this soft gluon region of phase space then factorizes as
\begin{equation}
\frac{1}{\sigma}\frac{d^3\sigma}{dD\, dC\, dz} = H_{2-\text{jet}}(Q^2)H_s(z,C)J_1(C,D)\otimes J_2(C,D)\otimes J_3(z,D)\otimes S_{123}(z,C,D)\,.
\end{equation}
Here, we have expressed the cross section in terms of the value of the $C$- and $D$-parameters, and the energy fraction of the gluon, $z$.  \Fig{fig:reg2softscales} displays the hierarchal scales of this factorization theorem.  The anomalous dimension of the new hard function $H_s$ appearing in this factorization theorem is presented in \App{app:softreg2anom}.

\section{Combining Factorization Theorems: Resummation for $D\ll C^2$}\label{sec:comb}

The forms of the factorization theorems for either the collinear gluon jet or the soft gluon jet in region 2 both followed from a refactorization of the factorization theorem of region 1.  That is, a consistent combination of the cross sections in regions 1 and 2 will result in a cross section for the region where $D\ll C^2$, with no constraint on the absolute scale of the $C$-parameter.  One way to do this combination is to match the cross sections of the different regions and then subtract their overlap.  This was effectively the approach taken in the calculations of the $D_2$ observable \cite{Larkoski:2013eya,Larkoski:2014gra} in the two-prong region of phase space in \Refs{Larkoski:2015kga,Larkoski:2017iuy,Larkoski:2017cqq} or the approach advocated in the general hierarchical jet analysis in \Ref{Pietrulewicz:2016nwo}.  With one-loop anomalous dimensions, such a combination of regions would correctly resum all large logarithms for $D\ll C^2$ to NLL accuracy.

However, for our purposes in this paper, there is a significantly simpler way in which to combine phase space regions 1 and 2.  We present this procedure which accomplishes approximate NLL accuracy, only missing logarithms from the collinear-soft function $C_s(C,D,z)$.  We can get away with approximate NLL accuracy for making plots because a complete calculation of the $D$-parameter requires including region 3 as well, but there currently exists no factorization theorem for this region, as will be discussed in the next section.

Recall the factorization theorem for region 1, when $D \ll C^2 \sim 1$:
\begin{equation}
\frac{1}{\sigma}\frac{d^3\sigma}{dD\, dx_1\, dx_2} = H_{3-\text{jet}}(x_1,x_2)J_1(x_1,D)J_2(x_2,D)J_3(2-x_1-x_2,D)S_{123}(x_1,x_2,D)\,.
\end{equation}
In region 2, because we still assume a hierarchy between the values of $D$ and $C$, the functions set by the measured value of the $D$-parameter (the jet functions and soft function) are identical to the calculation of the corresponding function in region 1.  In the evaluation of region 2, we have just taken the appropriate collinear or soft limit of the emission that sets the value of the $C$-parameter.  This requires refactorization of the hard function and refactorization of the soft function for collinear subjets.  In region 1, because $C\sim 1$, there is no hierarchy between the center-of-mass energy and the scale of the $C$-parameter.  However, in region 2, there is a hierarchy, and this is resummed by either the function $H_{1\to 2}$ or $H_s$, as appropriate to the collinear or soft limits, respectively.  In the following, we present a single formula that resums all logarithms present in the hard and jet functions of regions 1 and 2, and only misses those logarithms that arise from refactorization of the soft function.

To completely describe the region where $D\ll C^2$, we need to account for these different regions.  To our approximate NLL accuracy, we can do this in the following way.  We construct a new factorized form of the cross section:
\begin{equation}\label{eq:reg12tot}
\frac{1}{\sigma}\frac{d^3\sigma^{D\ll C^2}}{dD\, dx_1\, dx_2} = H(Q^2)H_{(2\to 3)}(x_1,x_2)J_1(x_1,D)J_2(x_2,D)J_3(2-x_1-x_2,D)S_{123}(x_1,x_2,D)\,.
\end{equation}
Here, $H(Q^2)$ is the hard function for $e^+e^-\to q\bar q$ production, which to lowest order is just 1 and has an anomalous dimension of 
\begin{equation}
\gamma_H = -2\frac{\alpha_s C_F}{\pi}\log\frac{\mu^2}{Q^2}-3\frac{\alpha_s C_F}{\pi}\,.
\end{equation}
Using the results presented in \App{app:reg1}, we then imply the anomalous dimension of the ``antenna function'' $H_{(2\to 3)}(x_1,x_2)$ by renormalization group invariance of the cross section to be
\begin{align}
\gamma_{H_{(2\to 3)}} &= -\frac{\alpha_sC_A}{\pi}\log\frac{6\mu^2}{(2-x_1-x_2)CQ^2}-\frac{\alpha_s}{\pi}\frac{11C_A-2 n_f}{6}\\
&
\hspace{1cm}
-\frac{\alpha_sC_A}{\pi}\log\frac{(x_1+x_2-1)^2}{x_1x_2}+2\frac{\alpha_s C_F}{\pi}\log(x_1+x_2-1)\,.\nonumber
\end{align}

By the renormalization group invariance of the cross section, we can fix the renormalization scale $\mu = Q$, the center-of-mass energy.  This enables us to consistently resum over the entire range when $D\ll C^2$.  Note that the natural scale of the hard function $H(Q^2)$ is set to $Q$ also: $\mu_H = Q$.  That is, there is no running of the hard function from its natural scale.  Therefore, with this prescription $H(Q^2) = 1$.  Then, one needs to appropriately set the natural scale of the antenna function $H_{(2\to 3)}(x_1,x_2)$ in order to resum all logarithms of the $C$-parameter.  When $C\sim 1$, we want the natural scale of $H_{(2\to 3)}(x_1,x_2)$ to be $\mu_{H_{(2\to 3)}}=Q$ so that it takes its tree-level value:
\begin{equation}
H^{(0)}_{(2\to 3)}(x_1,x_2) = \frac{\alpha_s C_F}{2\pi}\frac{x_1^2+x_2^2}{(1-x_1)(1-x_2)}\,.
\end{equation}
That is, when $C\sim 1$, the hard function $H_{(2\to 3)}$ has no large logarithms because we choose $\mu=Q$.  When $C^2 \ll 1$, the natural scale of this function should correspond to resumming the hierarchy between the center-of-mass scale $Q$ and the value of the $C$-parameter.  From the results presented in \App{app:reg2app}, this scale is
\begin{equation}\label{eq:newscalenll}
\mu_{H_{(2\to 3)}} = \sqrt{\frac{(2-x_1-x_2)C}{6}}Q\,.
\end{equation}
So, we just need to design a scale that interpolates between these two regimes, as a function of the value of the $C$-parameter.

Our prescription for doing this will be very simple.  We just multiply the natural scale of $H_{(2\to 3)}(x_1,x_2)$ by a factor so that it matches onto the hard scale $\mu_H = Q$ at the kinematic endpoint.  The maximum value of the $C$-parameter at ${\cal O}(\alpha_s)$ is $3/4$ at which point all phase space variables $x_i=2/3$.  At this kinematic endpoint, the scale of \Eq{eq:newscalenll} is therefore
\begin{equation}
\mu_{H_{(2\to 3)}}^\text{endpoint} =\frac{Q}{\sqrt{12}}\,.
\end{equation}
So, we rescale $\mu_{H_{(2\to 3)}}$ by a factor of $\sqrt{12}$:
\begin{equation}
\mu_{H_{(2\to 3)}} \to \sqrt{2(2-x_1-x_2)C}Q\,,
\end{equation}
and rearrange the anomalous dimension of this function appropriately:
\begin{align}
\gamma_{H_{(2\to 3)}} &= -\frac{\alpha_sC_A}{\pi}\log\frac{\mu^2}{2(2-x_1-x_2)CQ^2}-\frac{\alpha_s}{\pi}\frac{11C_A-2 n_f}{6}\\
&
\hspace{1cm}
-\frac{\alpha_sC_A}{\pi}\log\frac{12(x_1+x_2-1)^2}{x_1x_2}+2\frac{\alpha_s C_F}{\pi}\log(x_1+x_2-1)\,.\nonumber
\end{align}
We will use this form of the anomalous dimension to produce our results that we plot later.  This scale setting approach exactly reproduces the factorization theorems of region 1 and the soft jet of region 2, so it will successfully resum all of those logarithms.  It misses, however, the re-factorization of the soft function of the collinear jet of region 2, and therefore does not completely capture all logarithms of this region, which is why we refer to it as approximate NLL.

\section{Region 3 (Dijets): $D \sim C^2\ll 1$}\label{sec:reg3}

The third region defined by the simultaneous measurement of the $C$- and $D$-parameters is the most subtle.  In regions 1 and 2, because $D\ll C^2$, the emissions that set the $C$-parameter also set the event plane.  The emissions that set the $D$-parameter, however, had too low of a relative $k_T$ to affect the event plane, and so the $D$-parameter is additive.  This property is vital for the factorization of the cross section in \Eq{eq:reg12tot}, for example.  The convolutions over the value of the $D$-parameter are a consequence of this additivity, and so resummation in regions 1 and 2 is straightforward.

In region 3, where $D\sim C^2\ll 1$, the $D$-parameter is no longer additive.  Emissions that set $C$ also affect $D$, and vice-versa, so the event plane is not well-defined.  As such, this seems to suggest that additional emissions need to know about all previous emissions to set the event plane, and therefore the value of $D$.  This is somewhat of a similar situation to that of recoil-sensitive observables such as broadening \cite{Rakow:1981qn,Ellis:1986ig,Catani:1992jc}, and may suggest that resummation requires cross-talk between soft and jet functions in addition to just their measured values of $D$ \cite{Chiu:2012ir}.   At the very least, the resummation in this region will be non-standard.  So, we will approach its calculation from a different perspective, rather than attempting factorization of the cross section.

Let $p(C,D)$ be the joint probability distribution of the $C$- and $D$-parameters.  This is just the (normalized) double differential cross section.  Then, from the definition of conditional probability, this can be re-written as
\begin{equation}
p(C,D) = p(C)p(D|C)\,,
\end{equation}
where $p(C)$ is the probability distribution of the $C$-parameter and $p(D|C)$ is the conditional probability of $D$ given $C$.  Expressed in terms of differential cross sections, we have
\begin{equation}
\frac{d^2\sigma}{dC\, dD} = \frac{d\sigma}{dC}\frac{d\sigma(C)}{dD}\,,
\end{equation}
where $\sigma(C)$ denotes that $C$ is a parameter of the cross section, and not a random variable.  So far, this is an identity: as long as everything on both sides of this equation are calculated to the same accuracy, the equality is exact.  However, written in this form enables another approach that will allow us to make progress.

Our goal will be to calculate the double differential cross section of $C$ and $D$ in region 3 to NLL accuracy.  The cross section of the $C$-parameter $d\sigma/dC$ has been calculated to high perturbative accuracy both at fixed- and resummed-order \cite{Gehrmann-DeRidder:2007nzq,DelDuca:2016csb,DelDuca:2016ily,Hoang:2014wka}, and so getting it to NLL accuracy is no problem.  The conditional cross section, $d\sigma(C)/dD$, on the other hand, contains all of the subtleties of region 3, and (currently) cannot be resummed.  However, it can be calculated to fixed-order in region 3, which is what we will do here.  As the $D$-parameter is first non-zero at ${\cal O}(\alpha_s^2)$, we will calculate the conditional cross section to that order.  So, the expression for our resummed double differential cross section for the $C$- and $D$-parameters is
\begin{equation}\label{eq:nllreg3}
\frac{d^2\sigma^{\text{NLL}}}{dC\, dD} \approx \frac{d\sigma^{\text{NLL}}}{dC}\frac{d\sigma(C)^{\alpha_s^2}}{dD}\,.
\end{equation}
The superscripts denote the accuracy to which each factor is calculated.  The fixed-order conditional cross section is defined as
\begin{align}
\frac{d\sigma(C)^{\alpha_s^2}}{dD} \equiv \frac{\frac{d\sigma^{\alpha_s^2}}{dC\, dD}}{\frac{d\sigma^{\alpha_s}}{dC}}\,.
\end{align}
Note that this conditional cross section is technically ${\cal O}(\alpha_s)$, because it is formed from the ratio of a cross section at $\alpha_s^2$ to one at $\alpha_s$; i.e., the first non-trivial order for the numerator and denominator, respectively.  The lowest-order cross section of the $C$-parameter in the limit $C\ll 1$ is \cite{Catani:1998sf}
\begin{equation}\label{eq:singas}
\frac{1}{\sigma_0}\frac{d\sigma^{\alpha_s}}{dC} = 2\frac{\alpha_s C_F}{\pi}\frac{1}{C}\left[
\log\frac{6}{C}-\frac{3}{4}
\right]\,,
\end{equation}
where $\sigma_0$ is the Born-level cross section for $e^+e^-\to q\bar q$ scattering.  In the evaluation of \Eq{eq:nllreg3} for comparison to LEP data in later sections, we will include the effects of a running coupling in the cross section for the $C$-parameter.  The expression including the running coupling is provided in \App{app:reg3}.

This approach to constructing the cross section has been used to study observables that are formally infrared and collinear unsafe, and yet have finite cross sections when all-orders effects are accounted for.  Such observables are referred to as Sudakov safe \cite{Larkoski:2013paa,Larkoski:2015lea}.  In the analogy with the $C$- and $D$-parameters, $D$ would represent the unsafe observable, while $C$ would be a safe companion.  By measuring the safe companion, the conditional probability of the unsafe observable is calculable in fixed-order perturbation theory.  Then, one can marginalize over the safe companion and a finite result is obtained when a Sudakov factor is included to exponentially suppress the singular region of phase space.  Here, however, the $D$-parameter is itself infrared and collinear safe, so the analogy only goes so far.  Nevertheless, measuring the $C$-parameter was necessary to isolate this dijet region of phase space.

As written, \Eq{eq:nllreg3} is not restricted to region 3, where $D\sim C^2\ll 1$.  There is a non-trivial contribution from the region where $D\ll C^2\ll 1$, which is already covered by regions 1 and 2.  Therefore, we need to eliminate this contribution, which we will do by simple subtraction.  That is, to restrict to region 3, we just subtract the limit when $D\ll C^2$ from the conditional cross section:
\begin{equation}\label{eq:nllreg3sub}
\frac{d^2\sigma^{D\sim C^2}}{dC\, dD} \approx \frac{d\sigma^{\text{NLL}}}{dC}\left[\frac{d\sigma(C)^{\alpha_s^2}}{dD}-\frac{d\sigma(C)^{\alpha_s^2, D\ll C^2}}{dD}\right]\,.
\end{equation}
We calculate the subtracted double differential cross section of region 3 to $\alpha_s^2$ at single-logarithmic accuracy in \App{app:reg3}.  At single-logarithmic accuracy, we only need to consider the soft and collinear limit to $\alpha_s^2$, which simplifies the calculation.  We were unable to find exact expressions for the results in region 3, and we use numerical interpolation to perform the integrals over $C$ to determine the inclusive $D$-parameter cross section.

The running coupling here needs to be evaluated at the relative transverse momentum of the emissions.  This accomplishes a resummation of some of the collinear logarithms that arise at higher perturbative orders.  The exact scales and integrals are worked out in \App{app:reg3}, but to NLL accuracy, the running couplings can just be evaluated as:
\begin{equation}
\alpha_s^2 \equiv \alpha_s(CQ)\alpha_s\left(\sqrt{C}Q\right)\,.
\end{equation}
Here, $Q$ is the center-of-mass collision energy, and the one-loop running coupling is
\begin{equation}
\alpha_s(\mu) = \frac{\alpha_s(Q)}{1+\frac{\alpha_s(Q)}{2\pi}\beta_0\log\frac{\mu}{Q}}\,.
\end{equation}
$\beta_0$ is the one-loop coefficient of the QCD $\beta$-function
\begin{equation}
\beta_0 =\frac{11}{3}C_A -\frac{4}{3}T_R n_f\,.
\end{equation}

As mentioned above, this procedure does not strictly account for all logarithms that may be present at NLL accuracy.  However, we believe that it only misses potential contributions from soft emissions at angles comparable to the harder emissions that set the value of $C$ (or $D$).  The effect of collinear emissions off of the emissions that set the $C$- and $D$-parameters are included at NLL accuracy by evaluating the coupling $\alpha_s$ in \Eq{eq:nllreg3sub} at the relevant $k_T$ scale.   The resummed cross section of $C$ accounts for logarithmically-enhanced emissions off of the original quark--anti-quark dipole.  Soft emissions from dipoles formed from the emissions that set the value of $D$, on the other hand, resolve the structure of the event plane.  We do not expect them to, in general, be correctly accounted for in \Eq{eq:nllreg3}.  Nevertheless, \Eq{eq:nllreg3} enables a concrete prediction, formally accurate at least ${\cal O}(\alpha_s^2)$, and is systematically improvable by simply calculating the conditional probability to higher orders.  However, when logarithms of $D$ are comparable in size to the inverse of the coupling, the accuracy of the expression breaks down, and honest resummation of all terms is required.

The approximation used here is not equivalent to modified leading logarithmic accuracy (MLLA) \cite{Seymour:1997kj}, in which one just exponentiates those logarithms that arise at leading-order.  For additive observables like $C$-parameter or thrust, MLLA is a sensible approximation, because it is clear how to improve the approximation.  For the $D$-parameter in this region, it is not obvious how to improve MLLA to higher accuracy, because it is not a priori obvious how logarithms exponentiate at higher orders.  This is why we advocate for the conditional probability approach.  Further, the effect of multiple emissions setting the value of the $D$-parameter will only first contribute at NNLL order.  Emissions that set both the values of $C$ and $D$ at leading power must be non-strongly ordered in both energy and angle.  The leading-order contribution therefore scales like $\alpha_s^2 \log^2 D$, because there need to be two emissions to be out of the event plane and those two emissions must be soft and collinear to one of the jets.  If there was a third emission that contributed at leading power to $D$, it cannot be strongly-ordered with respect to the first two emissions, and therefore it contributes at order $\alpha_s^3 \log^2 D$, or NNLL.

\section{Combining Factorization Theorems of Regions 1, 2, and 3}\label{sec:combine}

With the double differential cross sections in regions 1, 2, and 3 established, we now want to combine them and integrate over the $C$-parameter to produce the cross section exclusive in $D$ only.  As established above, the double differential cross section of the $C$- and $D$-parameters in the limit where $D\ll 1$ is just the sum of the contributions from regions 1 and 2 and region 3:
\begin{equation}
\frac{d^2\sigma^{D\ll 1}}{dC\, dD} = \frac{d^2\sigma^{D\ll C^2}}{dC\, dD} + \frac{d^2\sigma^{D\sim C^2}}{dC\, dD}\,.
\end{equation}
In region 3, where $D\sim C^2$, we have already removed the overlapping contribution with regions 1 and 2 which is why the cross section is just a sum.  Now, we just integrate this over $C$.  As discussed in \Sec{sec:obs}, the range of the $C$-parameter when $D\ll 1$ is
\begin{equation}
C\in \left[\sqrt{\frac{4}{3}D},\frac{3}{4}\right]\,.
\end{equation}
This is thus the bounds of integration for the integral over $C$:
\begin{equation}
\frac{d\sigma}{dD} = \int_{\sqrt{\frac{4}{3}D}}^{3/4}dC\, \frac{d^2\sigma^{D\ll 1}}{dC\, dD}\,.
\end{equation}
With this integral, we are in some sense done, but it is useful to analyze it a bit further.

Let's first focus on the contributions from regions 1 and 2, where $D\ll C^2$: 
\begin{equation}
\frac{d\sigma^{D\ll C^2}}{dD} = \int_{\sqrt{\frac{4}{3}D}}^{3/4}dC\, \frac{d^2\sigma^{D\ll C^2}}{dC\, dD}\,.
\end{equation}
Formally, the cross section is only accurate in the region where $D\ll C^2$; however, the lower bound of the integral extends to the region in which $D\sim C^2$.  This is perfectly fine: this cross section nevertheless resums some of the large logarithms in this region, but does not capture all of them.  To get all logarithms of $D$ where $D\sim C^2$, we of course need to include the contribution from region 3.

The region 3 expression for the cross section is
\begin{equation}
\frac{d\sigma^{D\sim C^2}}{dD} = \int_{\sqrt{\frac{4}{3}D}}^{3/4}dC\, \frac{d^2\sigma^{D\sim C^2}}{dC\, dD}\,.
\end{equation}
As we have already explicitly subtracted the contribution from the region $D\ll C^2$, the upper bound of $C\leq 3/4$ strictly only adds power corrections in $D$.  So, we can safely take the upper bound in this region to $\infty$:
\begin{equation}
\frac{d\sigma^{D\sim C^2}}{dD} = \int_{\sqrt{\frac{4}{3}D}}^{\infty}dC\, \frac{d^2\sigma^{D\sim C^2}}{dC\, dD}\,.
\end{equation}
Further, let's write the double differential cross section in this region using conditional cross sections, as in the previous section:
\begin{equation}
\frac{1}{\sigma}\frac{d\sigma^{D\sim C^2}}{dD} = \frac{1}{\sigma}\int_{\sqrt{\frac{4}{3}D}}^{\infty}dC\, \frac{d\sigma}{dC}\,\frac{d\sigma^{D\sim C^2}(C)}{dD}=\int_{\sqrt{\frac{4}{3}D}}^{\infty}dC\, \left[\frac{\partial}{\partial C}\Sigma(C)\right] \frac{d\sigma^{D\sim C^2}(C)}{dD}\,.
\end{equation}
In the second equality, we have written the differential cross section of the $C$-parameter as the derivative of the cumulative distribution of the $C$-parameter, $\Sigma(C)$.  We can then integrate by parts:
\begin{align}
\frac{1}{\sigma}\frac{d\sigma^{D\sim C^2}}{dD} &=\int_{\sqrt{\frac{4}{3}D}}^{\infty}dC\, \left[\frac{\partial}{\partial C}\Sigma(C)\right] \frac{d\sigma^{D\sim C^2}(C)}{dD}\\
&=\left.\Sigma(C)\frac{d\sigma^{D\sim C^2}(C)}{dD}\right|_{\sqrt{\frac{4}{3}D}}^{\infty}-\int_{\sqrt{\frac{4}{3}D}}^{\infty}dC\, \Sigma(C) \left[\frac{\partial}{\partial C}\frac{d\sigma^{D\sim C^2}(C)}{dD}\right] \nonumber\\
&=- \Sigma\left(C = \sqrt{\frac{4}{3}D}\right)\frac{d\sigma^{D\sim C^2}\left(C=\sqrt{\frac{4}{3}D}\right)}{dD}-\int_{\sqrt{\frac{4}{3}D}}^{\infty}dC\, \Sigma(C) \left[\frac{\partial}{\partial C}\frac{d\sigma^{D\sim C^2}(C)}{dD}\right]\,.\nonumber
\end{align}
In writing this expression, we use the fact that the cross sections vanish at $C=\infty$.  Because our approach to this region calculates the cross section for $C$ to NLL accuracy, while the conditional cross section is only calculated to ${\cal O}(\alpha_s^2)$, it will be a bit easier to numerically evaluate the final equality, after integration by parts.  

\section{Non-Perturbative Effects}\label{sec:np}

Our analysis thus far has been restricted to perturbation theory, but we also need to determine the regime in which this analysis is valid.  In this section, we estimate the size and effect of corrections due to non-perturbative physics on the $D$-parameter distribution.  The separation of the cross section into the three regions enables a straightforward estimation of non-perturbative effects in each region.  We can then identify the region with the largest non-perturbative correction to specify the range of validity of our perturbative calculation.  The regions with the largest non-perturbative effects are regions 1 and 2, in which $D\ll C^2$, which we will justify.

In regions 1 and 2, non-perturbative corrections become important when the natural scales of functions in the factorized cross section approach the non-perturbative scale of QCD, $\Lambda_\text{QCD}$.  From the results of \App{app:reg1}, the function with the lowest scale is the soft function; either $S_{123}$ in region 1 and the soft jet of region 2 or $S_{12}$ for the collinear jet of region 2.  The soft scale $\mu_S$
\begin{equation}
\mu_S^2 = \frac{4 D^2 Q^2}{81 C^2}\,.
\end{equation}
Setting this equal to the QCD scale $\Lambda_\text{QCD}$, we find that the $D$-parameter cross section in this region is dominated by non-perturbative effects when
\begin{equation}
D_\text{NP}\simeq \frac{9}{2}C\frac{\Lambda_\text{QCD}}{Q}< \frac{27}{8}\frac{\Lambda_\text{QCD}}{Q}\,.
\end{equation}
The inequality on the right was found by setting the $C$-parameter to its largest value, $C<3/4$.  This linear scaling in $\Lambda_\text{QCD}$ of non-perturbative corrections to the $D$-parameter agrees with what was found in \Ref{Banfi:2001pb}.

In region 3, the scale of the $C$-parameter itself is comparable to that of the $D$-parameter.  Therefore, non-perturbative corrections to the $D$-parameter are potentially further suppressed by the value of the $C$-parameter.  Non-perturbative effects dominate when the lower bound of the integral over the $C$-parameter enters the non-perturbative regime.  This lower bound is $D < \frac{3}{4}C^2$, and so the non-perturbative value of the $D$-parameter is
\begin{equation}
D_\text{NP}\simeq\frac{3}{4}C_\text{NP}^2\,,
\end{equation}
where $C_\text{NP}$ is the value of the $C$-parameter at which it becomes non-perturbative.  This non-perturbative scale is linearly proportional to the QCD scale,
\begin{equation}
C_\text{NP}\simeq \frac{\Lambda_\text{QCD}}{Q}\,,
\end{equation}
and so the non-perturbative value of $D$ in region 3 is suppressed by two powers of the QCD scale:
\begin{equation}
D_\text{NP}\simeq\frac{3}{4}\frac{\Lambda_\text{QCD}^2}{Q^2}\,.
\end{equation}
Because the these corrections scale with energy $Q$ parametrically smaller than for regions 1 and 2, they are formally suppressed.

Thus, we only will consider non-perturbative corrections from regions 1 and 2.  At the $Z$-pole, $Q=m_Z$, the $D$-parameter remains perturbative as long as
\begin{equation}
D \gtrsim \frac{27}{8}\frac{\Lambda_\text{QCD}}{m_Z} \simeq 0.04\,.
\end{equation}
As we will show in \Sec{sec:lep}, the best data on the $D$-parameter from LEP has several bins at values of the $D$-parameter that are below this value.

\section{Comparison to {\tt EVENT2}}\label{sec:event2}

Before comparison of our resummed predictions for the $D$-parameter with LEP data, we will compare the fixed-order expansion to the output of the program {\tt EVENT2} \cite{Catani:1996vz}.  {\tt EVENT2} calculates the fixed-order cross section for $e^+e^-\to q\bar q g$ to next-to-leading order.  As such, it can predict the leading-order expression for the $D$-parameter, which is first non-zero at ${\cal O}(\alpha_s^2)$.  With a program like {\tt EERAD3} \cite{Ridder:2014wza} or {\tt NLOJET++} \cite{Nagy:1997yn}, which can calculate the fixed-order cross section for $e^+e^-\to q\bar q g$ to next-to-next-to-leading order, we could, in principle, predict the $D$-parameter at next-to-leading order (${\cal O}(\alpha_s^3)$).  However, it will be non-trivial enough to agree at leading-order, as our resummed result consists of multiple moving parts.  

The first thing we need to do is to extract the ${\cal O}(\alpha_s^2)$ cross section from our resummed predictions in regions 1, 2, and 3.  In region 3, numerical estimates for the single-logarithmic contribution are presented in \App{sec:cdreg3appcalc}.  The fixed-order expression in regions 1 and 2 can be found from the expressions for the anomalous dimensions in \App{app:reg1}.  The anomalous dimensions for the jet and soft functions is presented in \App{app:anomsumm} in Laplace space for the $D$-parameter.  With the initial condition that the tree-level functions are just 1 in Laplace space, we can solve the anomalous dimension equations and inverse Laplace transform to find the expressions in real space.  Then, we sum the jet and soft functions and multiply by the tree-level matrix element for $e^+e^-\to q\bar qg$ scattering to find the cross section differential in $D$ and the three-body phase space variables $x_1$ and $x_2$.  To find the cross section exclusively in $D$, we then just integrate over $x_1$ and $x_2$.

Solving the anomalous dimension equations and inverse Laplace transforming, the sum of the jet and soft functions at ${\cal O}(\alpha_s)$ is
\begin{align}
J_1^{(1)}(x_1,D)+J_2^{(1)}(x_2,D)+J_3^{(1)}(x_3,D)+S_{123}^{(1)}(x_1,x_2,D) &=-\frac{\alpha_s}{\pi}(2C_F+C_A)\frac{\log \frac{4D}{3C^2}}{D}\\
&
\hspace{-9cm}
-\frac{\alpha_s C_F}{\pi}\frac{1}{D}\log\frac{C^2}{36}+\frac{\alpha_s}{\pi}(C_A-C_F)\frac{1}{D}\log\frac{x_1x_2}{(x_1+x_2-1)^2} -\frac{\alpha_s}{12\pi}\frac{1}{D}(18C_F+11 C_A-2n_f)\,.
\nonumber
\end{align}
The superscript $(1)$ denotes that this is the result at ${\cal O}(\alpha_s)$.  Then, the full cross section triply differential in $D$, $x_1$, and $x_2$ is
\begin{align}\label{eq:reg12tripdiffalphas2}
\hspace{-0.3cm}
\frac{1}{\sigma_0}\frac{d^3\sigma^{(\alpha_s^2)}}{dx_1\, dx_2\, dD} &=H^{(0)}_{3-\text{jet}}(x_1,x_2) \left(
J_1^{(1)}(x_1,D)+J_2^{(1)}(x_2,D)+J_3^{(1)}(x_3,D)+S_{123}^{(1)}(x_1,x_2,D)
\right)\nonumber\\
&
=\frac{\alpha_s^2C_F}{2\pi^2}\frac{1}{D}\frac{x_1^2+x_2^2}{(1-x_1)(1-x_2)}\left[
-(2C_F+C_A)\log \frac{4D}{3C^2}
-C_F\log\frac{C^2}{36}\right.\\
&
\hspace{4cm}
\left.+(C_A-C_F)\log\frac{x_1x_2}{(x_1+x_2-1)^2} -\frac{18C_F+11 C_A-2n_f}{12}
\right]\nonumber\,.
\end{align}
Recall that the value of the $C$-parameter in this region is
\begin{equation}
C=6\frac{(1-x_1)(1-x_2)(x_1+x_2-1)}{x_1 x_2 (2-x_1-x_2)}\,.
\end{equation}

The cross section inclusive over $x_1$ and $x_2$ is found by integrating over them.  However, the integration only extends to the point when $C = \sqrt{\frac{4}{3}D}$, so the leading logarithmic term in \Eq{eq:reg12tripdiffalphas2} stays positive.  So, we need to integrate over $x_1$ and $x_2$ with this constraint.  Then in this region, we find
\begin{align}\label{reg12a2calc}
\frac{d\sigma^{(\alpha_s^2)}}{dD} &=\int_0^1dx_1 \int_0^1 dx_2\, \frac{d^3\sigma^{(\alpha_s^2)}}{dx_1\, dx_2\, dD}\\
&
\hspace{3cm}
\times \Theta(x_1+x_2-1)\,\Theta\left(
6\frac{(1-x_1)(1-x_2)(x_1+x_2-1)}{x_1 x_2 (2-x_1-x_2)} - \sqrt{\frac{4}{3}D}
\right)\nonumber\,.
\end{align}
This integral can be done numerically and combined with the result from region 3.

\subsection{Leading-Logarithmic Cross Section}\label{sec:ll}

When comparing to the limit in region 3, it will be useful to explicitly evaluate the leading-logarithmic cross section for the $D$-parameter in each color channel.  To do this, we take the soft and collinear limit of the $q\bar q g$ matrix element, and keep only the leading terms in the $x_1,x_2\to 1$ limit.  In doing this, we can rewrite the cross section in terms of the gluon's energy fraction $z$ and splitting angle $\theta^2$:
\begin{align}
\hspace{-0.3cm}
\frac{1}{\sigma_0}\frac{d^3\sigma^{(\alpha_s^2,\text{LL})}}{dz\, d\theta^2\, dD} &=2\frac{\alpha_s^2C_F}{\pi^2}\frac{1}{D}\frac{1}{z\theta^2}\left[
-(2C_F+C_A)\log \frac{4D}{3C^2}
-C_F\log\frac{C^2}{36}\right.\\
&
\hspace{4cm}
\left. -\frac{18C_F+11 C_A-2n_f}{12}
\right]\Theta\left(
\frac{3}{2}z\theta^2 - \sqrt{\frac{4}{3}D}
\right)\nonumber\,.
\end{align}
In this expression, we note that 
\begin{equation}
C=\frac{3}{2}z\theta^2\,.
\end{equation}
It's also useful to change variables to the ratio $y=D/C^2$, so the leading-logarithmic limit can be identified with $y\ll 1$.  The cross section then becomes
\begin{align}
\hspace{-0.3cm}
\frac{1}{\sigma_0}\frac{d^3\sigma^{(\alpha_s^2, \text{LL})}}{dy\, d\theta^2\, dD} &=\frac{\alpha_s^2C_F}{\pi^2}\frac{1}{D}\frac{1}{y\theta^2}\left[
-(2C_F+C_A)\log \frac{4y}{3}
-C_F\log\frac{D}{36y}\right.\\
&
\hspace{4cm}
\left. -\frac{18C_F+11 C_A-2n_f}{12}
\right]\Theta\left(
\frac{3}{4}-y
\right)\Theta\left(
\theta^2-\frac{2}{3}\sqrt{\frac{D}{y}}
\right)\nonumber\,.
\end{align}

To continue, we can integrate over $\theta^2\in[0,1]$ and find
\begin{align}
\hspace{-0.3cm}
\frac{1}{\sigma_0}\frac{d^2\sigma^{(\alpha_s^2, \text{LL})}}{dy\, dD} &=-\frac{\alpha_s^2C_F}{2\pi^2}\frac{\log\frac{4D}{9y}}{yD}\left[
-(2C_F+C_A)\log \frac{4y}{3}
-C_F\log\frac{D}{36y}\right.\\
&
\hspace{4cm}
\left. -\frac{18C_F+11 C_A-2n_f}{12}
\right]\Theta\left(
\frac{3}{4}-y
\right)\Theta\left(
y-\frac{4}{9}D
\right)\nonumber\,.
\end{align}
To isolate the leading-logarithmic cross section in each color channel, we can set all numerical factors in logarithms to 1 and remove subleading logarithmic terms.  The cross section then further reduces to
\begin{align}
\hspace{-0.3cm}
\frac{1}{\sigma_0}\frac{d^2\sigma^{(\alpha_s^2, \text{LL})}}{dy\, dD} &=\frac{\alpha_s^2C_F}{2\pi^2}\frac{\log\frac{D}{y}}{yD}\left[
C_F\left(2\log y+\log\frac{D}{y}\right)+C_A\log y
 -\frac{n_f}{6}
\right]\Theta\left(
1-y
\right)\Theta\left(
y-D
\right)\,.
\end{align}
To determine the cross section exclusive in only $D$, we just integrate over $y$.  This produces:
\begin{align}
\hspace{-0.3cm}
\frac{1}{\sigma_0}\frac{d\sigma^{(\alpha_s^2, \text{LL})}}{dD} &=-\frac{\alpha_s^2C_F}{\pi^2}\frac{1}{D}\left[
\frac{C_F}{3}\log^3 D+\frac{C_A}{12}\log^3 D-\frac{n_f}{24}\log^2 D
\right]\,.
\end{align}

\subsection{${\cal O}(\alpha_s^2)$ Cross Section and {\tt EVENT2}}

\begin{figure}[t]
\centering
\subfloat[]{\label{fig:ev2a2b0a2l2}
\includegraphics[width=0.45\linewidth]{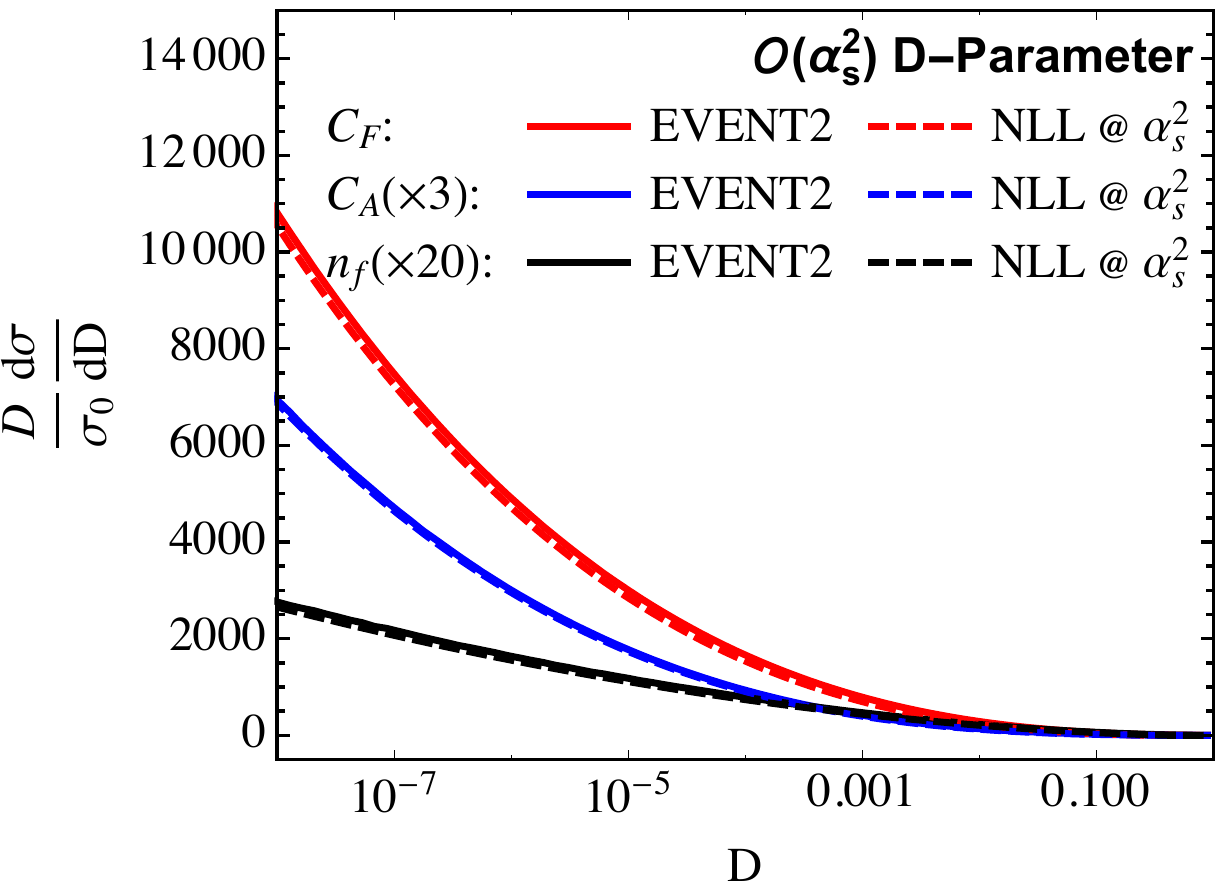}
}\quad
\subfloat[]{\label{fig:ev2a2b0a2l1}
\includegraphics[width=0.45\linewidth]{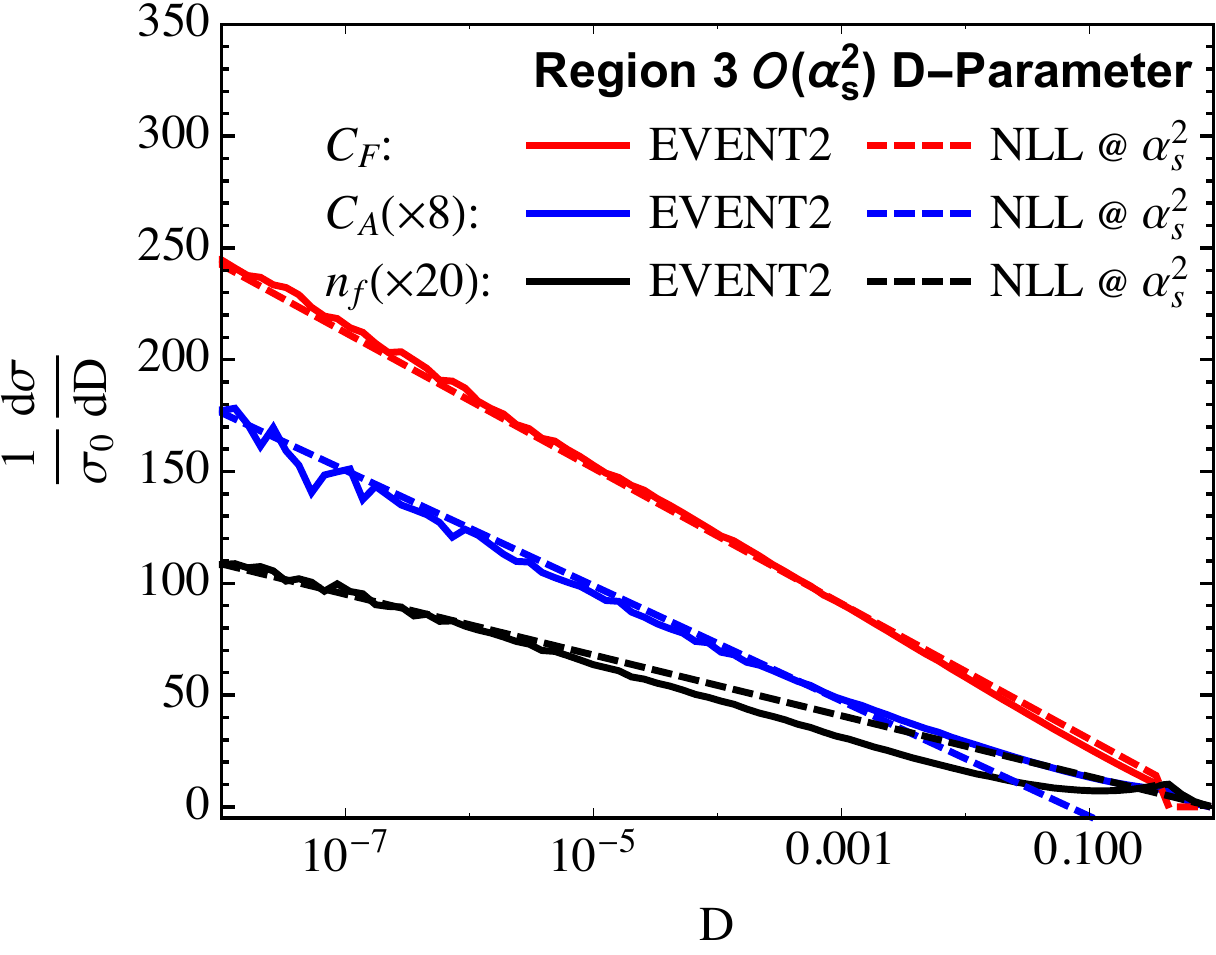}
}\\
\caption{
Comparison of analytic predictions for the distribution of the $D$-parameter at ${\cal O}(\alpha_s^2)$ through single-logarithmic accuracy (dashed) to {\tt EVENT2} (solid).  (a) The full logarithmic comparison exhibiting excellent agreement between analytics and {\tt EVENT2}.  (b) The comparison of the prediction in region 3 to {\tt EVENT2} with the contributions from regions 1 and 2 subtracted.}
\label{fig:event2comp} 
\end{figure}

With the numerical integral of \Eq{reg12a2calc} representing the contributions from regions 1 and 2, and the results from \App{sec:cdreg3appcalc} for region 3, we can validate our expression for the $D$-parameter cross section with {\tt EVENT2}.  The results of this comparison are shown in \Fig{fig:event2comp}.  Shown here are the differential cross sections plotted logarithmically in the value of $D$.  The individual contributions are separated into the three color channels at ${\cal O}(\alpha_s^2)$, $C_F^2$, $C_F C_A$, and $C_F n_f T_R$.  The Born-level cross section $\sigma_0$ has been divided out, as well as the appropriate color factor and an overall factor of $(\alpha_s/2\pi)^2$.  On the left in \Fig{fig:event2comp}, we compare the three different color channels as computed with {\tt EVENT2} and from our factorization theorems for the different regions (labeled as NLL).  Excellent agreement is observed, with only a relative constant offset between the {\tt EVENT2} results and our calculation.  This difference is beyond NLL accuracy, and so demonstrates consistency of this calculation.

The plot on the right of \Fig{fig:event2comp} focuses in on the contribution from region 3 specifically.  In this plot, we have subtracted the region 1--2 contribution from the {\tt EVENT2} distributions.  The result of this subtraction is a residual cross section that is linear in each color channel on this logarithmic plot.  This demonstrates that \Eq{reg12a2calc} does correctly describe the leading-logarithmic contributions that scale like $\alpha_s^2 \log^3 D/D$ and $\alpha_s^2 \log^2 D/D$.  The residual $\alpha_s^2\log D/D$ terms that remain are described by contributions from region 3.  As calculated in \App{sec:cdreg3appcalc}, the approximate logarithmic slopes of the different color channels from region 3 are:
\begin{align}
C_F:&\hspace{1cm }\approx-13.16\,,\\
C_A:&\hspace{1cm }\approx-1.4\,,\\
n_f T_R:& \hspace{1cm }\approx-0.31\,.
\end{align}
To evaluate these coefficients, we had to numerically integrate over the non-strongly-ordered soft and collinear region of phase space in each color channel.  The $C_F$ channel is significantly simple that all but one numerical integral could be done analytically, so its slope is known to high precision.  In the $n_f T_R$ channel, two numerical integrals must be done, but there is no strongly-ordered soft divergence which makes the matrix element regular over a large region of phase space.  So, the slope is also known well for this color channel.  This is in contrast to the $C_A$ color channel in which the strongly-ordered soft and collinear divergences in the matrix element make numerical integration significantly challenging.  Nevertheless, these slopes are in good agreement with the result of {\tt EVENT2}, given these caveats.  However, as \Fig{fig:event2comp} demonstrates, the $C_F$ channel dominates the cross section, so any uncertainties on numerical integration of the result from the $C_A$ channel will only have a small effect.

\section{Comparison to LEP Data}\label{sec:lep}

With our resummed predictions for the $D$-parameter in hand, in this section we compare to data.  We start in this section by first comparing our resummed predictions to fixed-order distributions to gain an understanding of how resummation is affecting the cross section.  We will compare our resummed results to both leading fixed order from {\tt EVENT2} as well as next-to-leading fixed order of the $D$-parameter from \Ref{Nagy:1997yn} which used the {\tt NLOJET++} program.  Our resummed prediction is a combination of the region 1, 2, and 3 predictions as discussed in earlier sections, and leading-order matching.  This is accomplished by simple additive matching:
\begin{equation}
\frac{d\sigma^\text{total}}{dD} = \frac{d\sigma^\text{resum}}{dD}+\frac{d\sigma^\text{LO}}{dD}-\frac{d\sigma^{\text{resum},\alpha_s^2}}{dD}\,.
\end{equation}
The leading-order cross section (``LO'') is the result from {\tt EVENT2}, the resummed cross section (``resum'') is the combination of regions 1, 2, and 3, and the fixed-order expansion of the resummed result (``$\text{resum},\alpha_s^2$'') was discussed in \Sec{sec:event2}.

\begin{figure}[t]
\centering
\subfloat[]{\label{fig:logresumfo}
\includegraphics[width=0.45\linewidth]{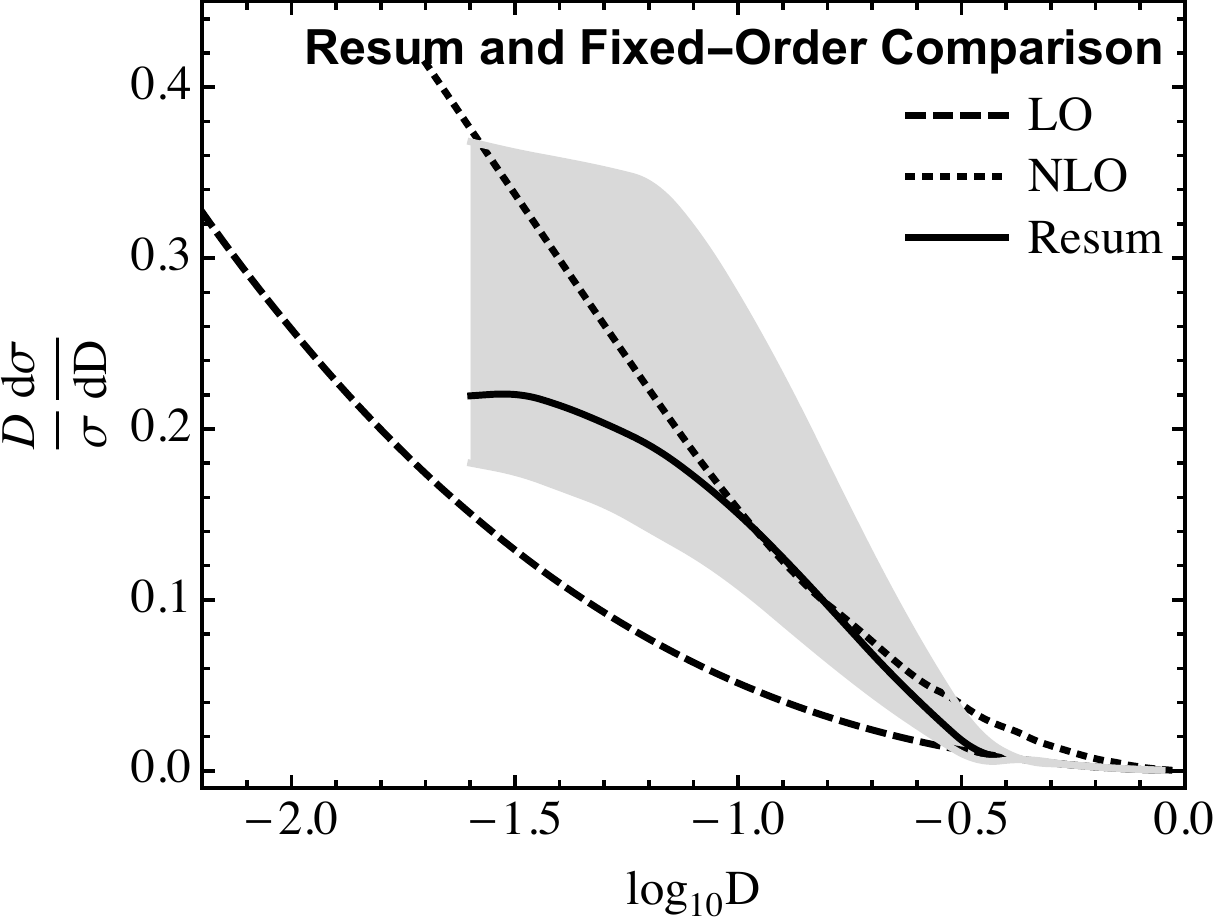}
}\quad
\subfloat[]{\label{fig:linresumfo}
\includegraphics[width=0.45\linewidth]{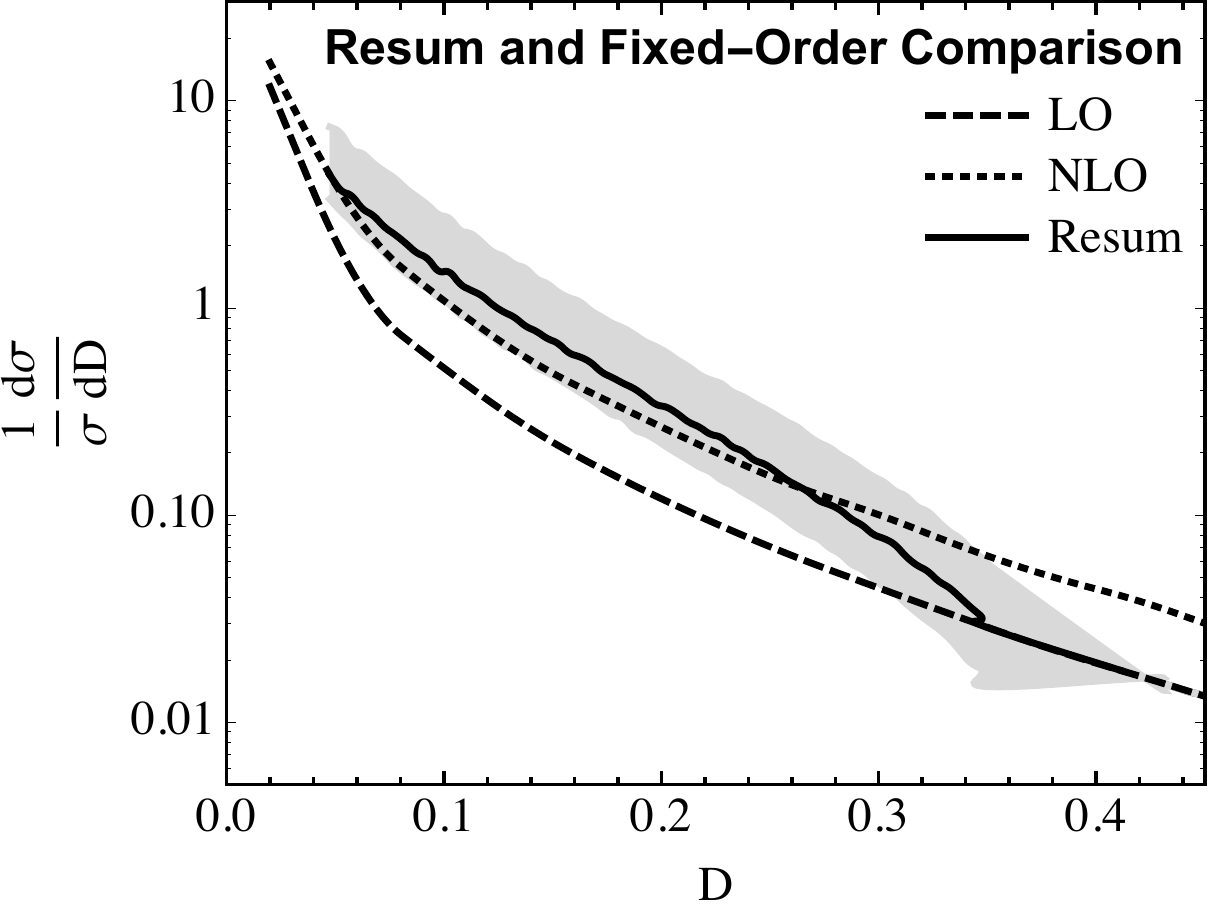}
}\\
\caption{
Comparison of our resummed prediction to leading and next-to-leading fixed order distributions on a logarithmic scale (left) and a linear scale (right).  The gray band represents an estimate of scale uncertainties in the resummed prediction.  The termination of the resummed prediction just below $D\simeq10^{-1.5}$ is due to the QCD Landau pole.
}
\label{fig:resumfocomp} 
\end{figure}

\Fig{fig:resumfocomp} compares these three cross sections plotted logarithmically (left) and linearly (right).  Our resummed predictions terminate when scales hit the QCD Landau pole to ensure that they remain in their region of validity.  As observed in \Refs{Nagy:1997yn,Campbell:1998nn}, the NLO $K$-factor is about a factor of 2, demonstrating that higher-order effects are very important for the $D$-parameter.  These plots demonstrate that this large NLO $K$-factor seems to be largely reproduced by resummation, and so is dominantly described by higher-order logarithms.  Resummation appears to become important below a value of about $D\lesssim 0.1$ where the cross section begins to turn over in the log-plot of \Fig{fig:logresumfo}.  The scale uncertainty band of the resummed cross section is estimated by varying scales in the expression for the cross section by a factor of 2.  There are many scales that exist in the resummed cross section, but the most sensitive scale is that of the soft function in the region 1--2 factorization theorem.  The scale uncertainty band in \Fig{fig:resumfocomp} is exclusively from varying the soft function scale up and down by a factor of 2.  We leave a detailed uncertainty analysis to future work.

It is also useful to discuss which phase space regions are relevant in these plots.  Recall that the upper limit on the value of $D$ in the resummation region is
\begin{equation}
D \leq \frac{3}{4}C^2\,,
\end{equation}
and with $C\leq 3/4$, the largest value $D$ can be is
\begin{equation}
D_{\max} = \frac{27}{64}\simeq 0.42\,.
\end{equation}
By contrast, the smallest value of $D$ for which our resummed distribution is perturbative corresponds to the location of the QCD scale or Landau pole relative to the $Z$ boson mass, at which 
\begin{equation}
D_{\min}\simeq 10^{-1.5}\simeq 0.03\,.
\end{equation}
Thus, our resummed prediction for the $D$-parameter ranges over only about one order of magnitude.  The factorization theorems of region 2, for which $D \ll C^2 \ll 1$ are numerically not very relevant because the largest value of the $C$-parameter in the resummation regime is $C_{\max} = 3/4$, while the minimum value is
\begin{equation}
C_{\min} = \sqrt{\frac{4}{3}D_{\min}} \simeq 0.2\,.
\end{equation}
That is, there just isn't phase space volume for the region 2 factorization theorems to contribute significantly.  The region 1 factorization theorem is relevant over the entire resummation regime of $D$, and the region 3 contribution becomes more important at smaller values of $D$.  Even at the smallest perturbative value of $D\simeq 0.03$, the region 1 contribution is still about a factor of three times larger than that of region 3.  

To be able to compare our prediction to data, we also need to include the effects of non-perturbative physics.  In \Sec{sec:np}, we demonstrated that regions 1 and 2, where $D\ll C^2$, has the largest non-perturbative corrections, and the $D$-parameter is additive in this region.   The average size of non-perturbative corrections to the $D$-parameter in regions 1 and 2 are
\begin{equation}
D_\text{NP}\simeq \frac{27}{8}\frac{\Lambda_\text{QCD}}{m_Z}\,.
\end{equation}
Because $D$ is additive in these regions, the dominant non-perturbative effects can be incorporated by simply translating the argument of the perturbative cross section in region 1 \cite{,Korchemsky:1994is,Dokshitzer:1995zt,Dokshitzer:1995qm,Dokshitzer:1997ew}:
\begin{equation}
\sigma^\text{reg.~1,full}(D) = \sigma^\text{reg.~1,pert.}\left(D-\frac{27}{8}\frac{\Lambda_\text{QCD}}{m_Z}\right)\,.
\end{equation}
Non-perturbative corrections can be included in a more detailed and quantitative way with shape functions \cite{Korchemsky:1999kt,Korchemsky:2000kp}, though such an analysis is beyond the scope of this paper.  All we will do to include the non-perturbative effects here is this translation of the regions 1--2 cross section argument.  In the following we set $\Lambda_\text{QCD}=1$ GeV so that $D_\text{NP}\simeq 0.038$.

\begin{figure}[t]
\centering
\subfloat[]{\label{fig:logdata}
\includegraphics[width=0.45\linewidth]{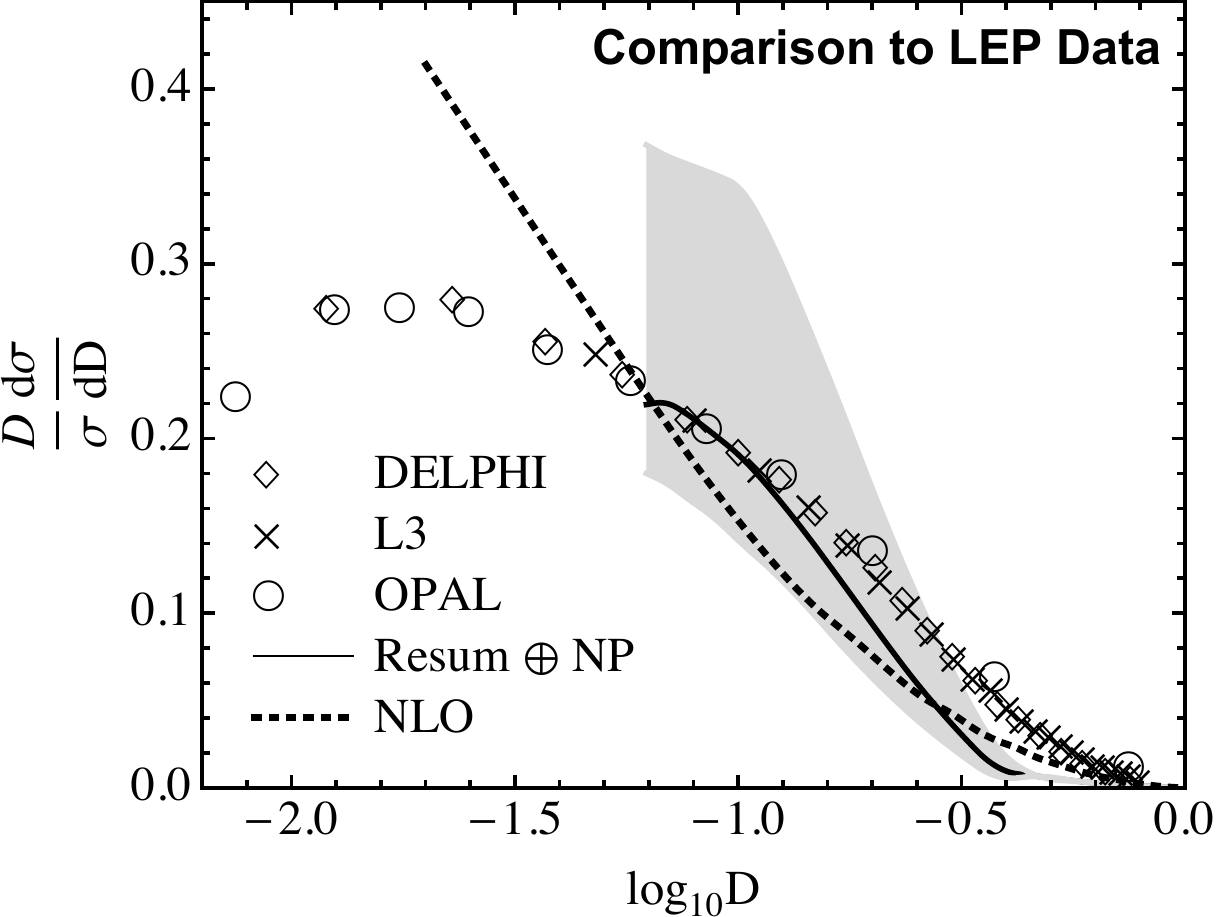}
}\quad
\subfloat[]{\label{fig:lindata}
\includegraphics[width=0.45\linewidth]{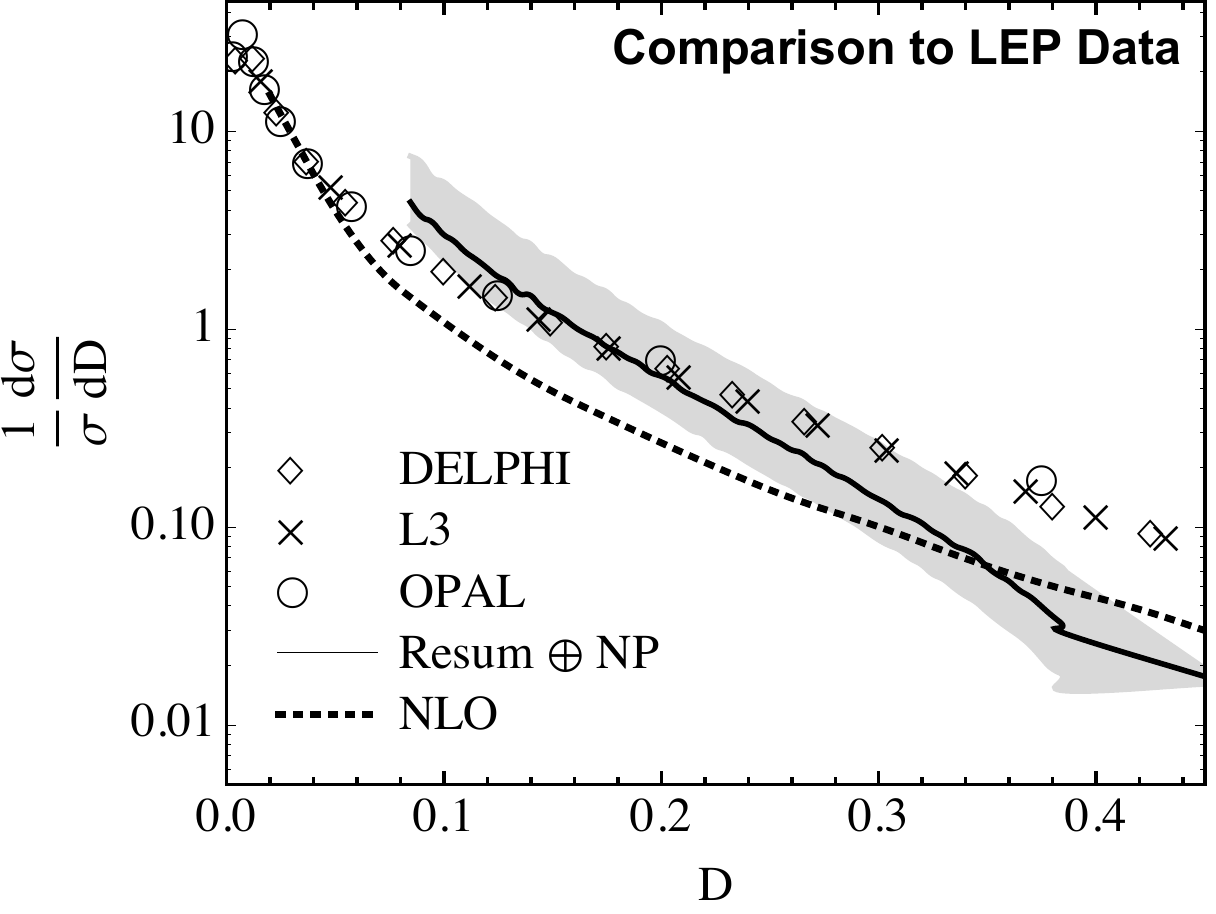}
}\\
\caption{
Comparison of our resummed prediction with non-perturbative corrections and the next-to-leading fixed order prediction to LEP data on a logarithmic scale (left) and a linear scale (right).  DELPHI represents data from \Ref{Abreu:1996na}, L3 represents data from \Ref{Achard:2004sv}, and OPAL represents data from \Ref{Abbiendi:2004qz}.  Uncertainties in the data are comparable in size to the plot markers.  The termination of the resummed prediction just below $D\simeq 0.1$ is where the non-perturbative corrections become large.
}
\label{fig:resumdatacomp} 
\end{figure}

The most precise measurements of the $D$-parameter were performed at LEP at a center-of-mass energy of the $Z$ boson mass \cite{Abreu:1996na,Achard:2004sv,Abbiendi:2004qz}.  While the $D$-parameter was measured at several other energies, we restrict our comparisons to the $Z$ pole to be able to make quantitative comparative statements.  This LEP data is compared to our resummed, leading-order matched, and including the non-perturbative shift in \Fig{fig:resumdatacomp}.  Good qualitative agreement is observed between our prediction and data, demonstrating that resummation is important for describing the data, especially in the turn-over observed in the log-plot in \Fig{fig:logdata} in data around $D\lesssim 0.1$.  Our resummed prediction significantly undershoots the data at large values of $D$, which we ascribe to a lack of higher-fixed order corrections.  We only match our resummed prediction to leading order, and so we miss the large $K$-factor between leading and next-to-leading order.  These plots also show that next-to-leading order predictions do indeed improve agreement with data over leading-order, at large values of $D$.  A complete matching to next-to-leading order that properly resums all logarithms requires next-to-next-to leading order resummation, which we leave to future work.

Fixed-order studies of the $D$-parameter \cite{Nagy:1997yn,Campbell:1998nn} noted that to match next-to-leading order predictions to LEP data requires an excessively small choice of renormalization scale.  These studies, however, did not include any estimates of non-perturbative corrections, which was first done in \Ref{Banfi:2001pb}.  The very small renormalization scale in the fixed-order studies is a proxy for these non-perturbative corrections.  However, as our analysis in this paper and \Fig{fig:resumdatacomp} demonstrates, these non-perturbative corrections are necessary for a quantitative description of data, in addition to resummation.  The $D$-parameter is sensitive to infrared scales in a very different manner then familiar observables like thrust, and so a complete quantitative description requires the inclusion of fixed-order, resummation, and non-perturbative physics.

\section{Conclusions}\label{sec:concs}

Resummation of the $D$-parameter throughout its phase space requires techniques that were only recently developed in the context of jet substructure for LHC applications.  Unlike familiar observables like thrust, the value of the $D$-parameter doesn't isolate individual phase space configurations as it is only sensitive to the aplanarity of the final state hadrons in $e^+e^-$ collision events.  To separate phase space configurations, we additionally measure the $C$-parameter on the final state, and different hierarchical relationships between $D$ and $C$ correspond to different phase space boundaries.  One of these regions, the near-to-planar three jet configuration, has been studied before, but the two other regions are novel.  We presented all-orders factorization theorems or systematically-improvable calculations in each region and compared resummed and matched predictions including estimates of non-perturbative effects to precision data measured at LEP.

This work is just the beginning for analytically understanding complicated event shapes measured at LEP.  A precision program at the level of thrust or $C$-parameter for observables like the $D$-parameter is not likely any time soon, but because $D$ probes significantly exotic phase space configurations, its uncertainties, both experimental and theoretical, are unique.  Therefore, predictions of the $D$-parameter at any accuracy are a detailed cross check on the consistency of strong coupling values, for example.  With improved predictions for the $D$-parameter, they might feed into Monte Carlo tunings.  Four-jet observables at $e^+e^-$ colliders are the first that are directly sensitive to the self-coupling of the gluon and non-trivial color flow, and therefore are used to set parameters like color reconnections within Monte Carlo simulations.

In the regions where the $D$-parameter is additive, we are able to derive all-orders factorization theorems for which each function is just convolved with one another.  When $D\sim C^2$, emissions that set the $D$-parameter therefore also set the event plane, and this backreaction is responsible for the $D$-parameter to lose additivity and prohibit any standard factorization.  Related issues with factorization breakdown in endpoint regions of phase space have also been observed in studies of observables sensitive to multi-prong jet substructure \cite{Larkoski:2015kga,Larkoski:2017cqq,Napoletano:2018ohv}.  Nevertheless, this does not imply that factorization is impossible.  With motivation from several examples, factorization of multi-differential cross sections away from strongly-ordered phase space boundaries would be a major advance and enable precision studies of a large class of observables relevant at both LEP and LHC.

The analysis of this paper also demonstrates that we continue to learn new ways of calculating in perturbative QCD that can be applied to old problems.  LEP left a legacy of numerous precision event shapes, of which only a very few are understood at sufficient accuracy to be used for extracting $\alpha_s$, for example.  We hope that this rich data inspires more studies and enables more advances in analytically studying QCD in extreme phase space regions.

\acknowledgments

A.L.~thanks Andrea Banfi for originally suggesting this problem, for collaboration in early stages of this work, and comments on the manuscript.  A.L.~also thanks Ian Moult and Duff Neill for comments on the manuscript, and Wouter Waalewijn and Frank Tackmann for addressing an error in the original description of the region 2 factorization theorems.

\appendix

\section{Region 1 Calculations}\label{app:reg1}

\subsection{Operator Definitions of Functions in Factorization Theorem}

In this appendix, we provide the SCET operator definitions for the jet and soft functions that appear in the factorization theorem of region 1, \Eq{eq:factreg1}.  For a quark jet function $J_q(x_q,D)$, it is defined as the following matrix element of SCET operators:
\begin{align}
J_q(x_q,D) = \frac{(2\pi)^3}{C_F} \text{tr}\,\langle0|
\frac{\slash \!\!\! \bar n}{2}\chi_n(0)\delta(x_q Q-\bar n\cdot {\cal P})\delta^{(2)}(\vec {\cal P}_\perp)\delta(D-\hat {\bf D})
\bar\chi_n(0)|0\rangle\,.
\end{align}
Here, we assume that the jet direction is defined by the light-like vector $n$, the opposite direction is defined by $\bar n$ for which $n\cdot \bar n =2$.  In this expression, ${\cal P}$ is the momentum measurement operator which, for a collinear jet, has zero perpendicular component $\vec {\cal P}_\perp$, and whose large light-cone momentum component $\bar n\cdot {\cal P}$ is set equal to the center-of-mass energy $Q$ times the energy fraction $x_q$.  $\chi(x)$ is the gauge-invariant collinear quark field defined as
\begin{equation}
\chi_n(x) = \delta(\omega-\bar n\cdot {\cal P}) W_n^\dagger(x)\xi_n(x)\,.
\end{equation}
The $\delta$-function in this expression simply sets the large light-cone component of momentum to be $\omega$, $\xi_n(x)$ is the $n$-collinear quark field in SCET that transforms in the fundamental representation of SU(3) color, and $W_n(x)$ is a collinear Wilson line that ensures gauge invariance:
\begin{equation}
W_n(x) = \sum_\text{perms} \exp\left[
-\frac{g}{\bar n \cdot {\cal P}}\bar n\cdot A_n(x)
\right]\,.
\end{equation}
The sum runs over all permutations of exponential factors, $g$ is the QCD coupling, and $A_n(x)$ is the Lie algebra-valued $n$-collinear gluon field.  Finally, the operator $\hat {\bf D}$ measures the $D$-parameter on the collinear quark field, as defined by the leading-power expression for the $D$-parameter in the collinear limit, \Eq{eq:dcollreg1}.

The gluon jet function, $J_g(x_g,D)$, is defined similarly, as
\begin{align}
J_g(x_g,D) = \frac{(2\pi)^3}{C_A} \text{tr}\,\langle0|{\cal B}_{n,\perp}^\mu
\delta(x_g Q-\bar n\cdot {\cal P})\delta^{(2)}(\vec {\cal P}_\perp)\delta(D-\hat {\bf D})
{\cal B}_{n,\perp\,\mu}^\dagger|0\rangle\,.
\end{align}
Now, the gauge-invariant collinear gluon field is
\begin{equation}
{\cal B}_{n,\perp}^\mu = \frac{1}{g}\delta(\omega-\bar n\cdot {\cal P}) W_n^\dagger(x)\, iD_\perp^\mu\,
W_n(x)\,.
\end{equation}
The covariant derivative $D_\perp^\mu$ is composed of momentum operators and gluon fields as
\begin{equation}
iD_\perp^\mu = {\cal P}_\perp^\mu + g A_{n\perp}^\mu\,,
\end{equation}
where the $\perp$ subscript denotes that the momentum and fields only have non-zero components transverse to the $n$ direction; that is, describing physical, transverse polarized gluons.

The soft function that appears in the factorization theorem is defined as a matrix element of soft Wilson lines:
\begin{align}
S_{123}(x_1,x_2,D) = \langle 0 |\text{T}(S_1 S_2 S_3)\delta(D-\hat{\bf D})\bar{\text{T}}(S_1S_2 S_2)|0\rangle\,.
\end{align}
The soft Wilson line in the $i$ direction is
\begin{equation}
S_i = {\bf P}\exp\left[
ig\int_0^\infty ds\, n_i\cdot A(x+sn_i)
\right]\,,
\end{equation}
where $n_i$ is the light-like direction that defines $i$.  {\bf P} is the path-ordering operator and the color representation in which the soft gluon field $A(x)$ is evaluated is implicit, as defined by $i$'s particle type.  T and $\bar{\text T}$ are the time ordering and anti-ordering operators and the operator $\hat {\bf D}$ measures the value of the $D$-parameter in the soft limit, \Eq{eq:dsoftreg1}.  Note that the Wilson line is outgoing, as all colored particles are produced in the final state from the hard collision.  Finally, the dependence of the soft function on the energy fractions $x_1$ and $x_2$ is really a shorthand for dependence of the soft function on the angles between the Wilson lines, as soft emissions cannot resolve the energy of hard particles.

\subsection{Hard Function}

The leading-order hard function for the process $e^+e^-\to q\bar qg$ is just the tree-level matrix element:
\begin{equation}
H^{(0)}(x_1,x_2) = \frac{\alpha_s}{2\pi}C_F\frac{x_1^2+x_2^2}{(1-x_1)(1-x_2)}\,.
\end{equation}
Its one-loop anomalous dimension is \cite{Ellis:2010rwa}
\begin{align}
\gamma_H &= -\frac{\alpha_s}{\pi}(C_A+2C_F)\log\frac{\mu^2}{Q^2}-3\frac{\alpha_s C_F}{\pi}-\frac{\alpha_s}{\pi}\frac{11C_A-2 n_f}{6}\\
&
\hspace{1cm}
+\frac{\alpha_sC_A}{\pi}\log(1-x_2)+\frac{\alpha_sC_A}{\pi}\log(1-x_1)-\frac{\alpha_s}{\pi}(C_A-2C_F)\log(x_1+x_2-1)\nonumber\,.
\end{align}
In terms of the $C$-parameter, this can be expressed in the suggestive form as
\begin{align}
\gamma_H &= -2\frac{\alpha_sC_F}{\pi}\log\frac{\mu^2}{Q^2}-3\frac{\alpha_s C_F}{\pi}-\frac{\alpha_sC_A}{\pi}\log\frac{6\mu^2}{(2-x_1-x_2)CQ^2}-\frac{\alpha_s}{\pi}\frac{11C_A-2 n_f}{6}\\
&
\hspace{1cm}
-\frac{\alpha_sC_A}{\pi}\log\frac{x_1+x_2-1}{x_1x_2}-\frac{\alpha_s}{\pi}(C_A-2C_F)\log(x_1+x_2-1)\nonumber\,.
\end{align}

\subsection{Jet Functions}

\subsubsection{Quark Jet}

The phase space and matrix element for collinear emissions from a quark is
\begin{align}
\int \frac{d^dk}{(2\pi)^d}\delta(k^2)|{\cal M}(k)|^2&=\frac{\alpha_sC_F}{2\pi} \frac{(4\pi)^\epsilon}{\pi^{1/2}\Gamma(1/2-\epsilon)}\left(
\frac{4\mu^2}{x_i^2 Q^2}
\right)^\epsilon\int_0^1dz\int_0^\infty d\theta^2\int_0^\pi d\phi\,\sin^{-2\epsilon}\phi\\
&
\hspace{2cm}
\times (\theta^2)^{-1-\epsilon}z^{-2\epsilon}(1-z)^{-2\epsilon}\left(
\frac{1+(1-z)^2}{z}-\epsilon z
\right)\,.\nonumber
\end{align}
The $D$-parameter jet function is then
\begin{align}
J_i^q(x_i,D)&=\frac{\alpha_sC_F}{2\pi} \frac{(4\pi)^\epsilon}{\pi^{1/2}\Gamma(1/2-\epsilon)}\left(
\frac{4\mu^2}{x_i^2 Q^2}
\right)^\epsilon\int_0^1dz\int_0^\infty d\theta^2\int_0^\pi d\phi\,\sin^{-2\epsilon}\phi\\
&
\hspace{1cm}
\times (\theta^2)^{-1-\epsilon}z^{-2\epsilon}(1-z)^{-2\epsilon}\left(
\frac{1+(1-z)^2}{z}-\epsilon z
\right)\delta\left(
D-\frac{9}{2}Cx_i z(1-z)\theta^2 \sin^2\phi
\right)\nonumber\\
&=\frac{\alpha_sC_F}{2\pi} \frac{(4\pi)^\epsilon\pi^{1/2}}{\Gamma(1/2-\epsilon)}\left(
\frac{18C\mu^2}{x_i Q^2}
\right)^\epsilon\frac{1}{D^{1+\epsilon}}\int_0^1dz\, z^{-1-\epsilon}(1-z)^{-\epsilon}\left(
1+(1-z)^2-\epsilon z^2
\right)\,. \nonumber
\end{align}
Using the $+$-function expansion
\begin{equation}
z^{-1-\epsilon}=-\frac{1}{\epsilon}\delta(z)+\left(
\frac{1}{z}
\right)_+-\epsilon\left(
\frac{\log\, z}{z}
\right)_+ +\cdots\,,
\end{equation}
we find
\begin{align}
J_i(x_i,D)&=\frac{\alpha_sC_F}{2\pi} \frac{(4\pi)^\epsilon\pi^{1/2}}{\Gamma(1/2-\epsilon)}\left(
\frac{18C\mu^2}{x_i Q^2}
\right)^\epsilon\frac{1}{D^{1+\epsilon}}\int_0^1dz\, z^{-1-\epsilon}(1-z)^{-\epsilon}\left(
1+(1-z)^2-\epsilon z^2
\right) \\
&=\frac{\alpha_sC_F}{2\pi} \left(
\frac{18C\mu^2}{x_i Q^2}
\right)^\epsilon\frac{1}{D^{1+\epsilon}}\left[
-\frac{2}{\epsilon}+4\log 2-\frac{3}{2}
\right]\,.\nonumber
\end{align}
We only expand to ${\cal O}(\epsilon^{-1})$ to extract anomalous dimensions.

Laplace transforming the $D$-parameter, we find
\begin{align}
J_i(x_i,\tilde D)&=\frac{\alpha_sC_F}{\pi} \left[
\frac{1}{\epsilon^2}+\frac{1}{\epsilon}\log \frac{9C \tilde D\mu^2}{2x_i Q^2}+\frac{3}{4\epsilon}
\right]\,.\nonumber
\end{align}
Its anomalous dimension is therefore
\begin{equation}
\gamma_{q_i} = 2\frac{\alpha_sC_F}{\pi} \log \frac{9C \tilde D\mu^2}{2x_i Q^2}+\frac{3}{2}\frac{\alpha_sC_F}{\pi}\,.
\end{equation}

\subsubsection{Gluon Jet}

The jet function for a gluon jet on which the $D$-parameter is measured is
\begin{align}
J_i^q(x_i,D)&=\frac{\alpha_s}{2\pi} \frac{(4\pi)^\epsilon}{\pi^{1/2}\Gamma(1/2-\epsilon)}\left(
\frac{4\mu^2}{x_i^2 Q^2}
\right)^\epsilon\int_0^1dz\int_0^\infty d\theta^2\, (\theta^2)^{-1-\epsilon}\int_0^\pi d\phi\,\sin^{-2\epsilon}\phi\\
&
\hspace{0cm}
\times z^{-2\epsilon}(1-z)^{-2\epsilon}\left[
C_A\left(
\frac{1}{z}+\frac{1}{1-z}+z(1-z)-2
\right)+\frac{n_f}{2}\left(
1-\frac{2}{1-\epsilon}z(1-z)
\right)
\right]\nonumber\\
&
\hspace{0cm}
\times
\delta\left(
D-\frac{9}{2}Cx_i z(1-z)\theta^2 \sin^2\phi
\right)\nonumber\\
&
=\frac{\alpha_s}{2\pi} \frac{(4\pi)^\epsilon \pi^{1/2}}{\Gamma(1/2-\epsilon)}\left(
\frac{18 C\mu^2}{x_i Q^2}
\right)^\epsilon\frac{1}{D^{1+\epsilon}}\int_0^1dz\, z^{-\epsilon}(1-z)^{-\epsilon}\nonumber\\
&
\hspace{2cm}
\times \left[
C_A\left(
\frac{1}{z}+\frac{1}{1-z}+z(1-z)-2
\right)+\frac{n_f}{2}\left(
1-\frac{2}{1-\epsilon}z(1-z)
\right)
\right]\nonumber\,.
\end{align}
Using the $+$-function expansion, we find
\begin{align}
J_i^q(x_i,D)=\frac{\alpha_s}{2\pi} \frac{(4\pi)^\epsilon \pi^{1/2}}{\Gamma(1/2-\epsilon)}\left(
\frac{18 C\mu^2}{x_i Q^2}
\right)^\epsilon\frac{1}{D^{1+\epsilon}}\left(
-\frac{2}{\epsilon}C_A-\frac{11C_A-2n_f}{6}
\right)\,.
\end{align}

Laplace transforming the $D$-parameter, we find
\begin{equation}
J_i^q(x_i,\tilde D)=\frac{\alpha_s}{2\pi} \left(
\frac{2}{\epsilon^2}C_A+\frac{2}{\epsilon}\log\frac{9 C \tilde D\mu^2}{2x_i Q^2}+\frac{1}{\epsilon}\frac{11C_A-2n_f}{6}
\right)\,.
\end{equation}
Its anomalous dimension is therefore
\begin{equation}
\gamma_{g_i} = 2\frac{\alpha_s C_A}{\pi} \log\frac{9 C \tilde D\mu^2}{2x_i Q^2}+\frac{\alpha_s}{\pi}\frac{11C_A-2n_f}{6}\,.
\end{equation}

\subsection{Soft Function}\label{sec:sfreg1}

The one-loop calculation of the soft function for an emission off of jets $i$ and $j$ for the $D$-parameter in this region can be written as
\begin{align}
S_{ij}(D)&=-\frac{\alpha_s}{\pi}\mu^{2\epsilon}{\bf T}_i\cdot {\bf T}_j \frac{(4\pi)^\epsilon}{\pi^{1/2}\Gamma(1/2-\epsilon)}\int_0^\infty dk_\perp\, (k_\perp)^{1-2\epsilon}\int_{-\infty}^\infty d\eta\int_0^\pi d\phi\,\sin^{-2\epsilon}\phi\\
&
\hspace{7cm}
\times \frac{n_i\cdot n_j}{(n_i\cdot k)(n_j\cdot k)}\,\delta\left(
D-9C\frac{k_\perp}{Q}\frac{\sin^2\phi}{\cosh\eta}
\right)\,.\nonumber
\end{align}
We can choose coordinates such that
\begin{equation}
n_i\cdot k = k^+ = k_\perp e^{-\eta}\,.
\end{equation}
Then, $n_j\cdot k$ can be expressed as
\begin{align}
n_j\cdot k &= \frac{n_i\cdot n_j}{2}k^-+\frac{\bar n_i\cdot n_j}{2}k^+ - \sin\theta_{ij}k_\perp \cos\phi\\
&=\frac{n_i\cdot n_j}{2}k_\perp \left(
e^\eta+\cot^2\frac{\theta_{ij}}{2}e^{-\eta}-2\cot \frac{\theta_{ij}}{2}\cos\phi
\right)
\nonumber
\end{align}
The soft function is then
\begin{align}
S_{ij}(D)&=-2\frac{\alpha_s}{\pi}\mu^{2\epsilon}{\bf T}_i\cdot {\bf T}_j \frac{(4\pi)^\epsilon}{\pi^{1/2}\Gamma(1/2-\epsilon)}\int_0^\infty dk_\perp\, (k_\perp)^{-1-2\epsilon}\int_{-\infty}^\infty d\eta\int_0^\pi d\phi\,\sin^{-2\epsilon}\phi\\
&
\hspace{3cm}
\times \frac{1}{1+\cot^2\frac{\theta_{ij}}{2}e^{-2\eta}-2e^{-\eta}\cot \frac{\theta_{ij}}{2}\cos\phi}\,\delta\left(
D-9C\frac{k_\perp}{Q}\frac{\sin^2\phi}{\cosh\eta}
\right)\nonumber\\
&=-2\frac{\alpha_s}{\pi}{\bf T}_i\cdot {\bf T}_j \frac{(4\pi)^\epsilon}{\pi^{1/2}\Gamma(1/2-\epsilon)}\left(
\frac{81 C^2\mu^2}{Q^2}
\right)^\epsilon \frac{1}{D^{1+2\epsilon}}\int_{-\infty}^\infty d\eta\int_{-1}^1 dy\,(1-y^2)^{-1/2+\epsilon}\nonumber\\
&
\hspace{3cm}
\times \frac{\cosh^{-2\epsilon}\eta}{1+\cot^2\frac{\theta_{ij}}{2}e^{-2\eta}-2e^{-\eta}\cot \frac{\theta_{ij}}{2}y}\,.
\nonumber
\end{align}

The remaining integrals can be done with appropriate changes of variables.  With $x=e^{-\eta}$, the integrals become
\begin{align}
&\int_{-\infty}^\infty d\eta\int_{-1}^1 dy\,(1-y^2)^{-1/2+\epsilon} \frac{\cosh^{-2\epsilon}\eta}{1+\cot^2\frac{\theta_{ij}}{2}e^{-2\eta}-2e^{-\eta}\cot \frac{\theta_{ij}}{2}y}\\
&=2^{2\epsilon}\int_0^\infty dx\, x^{-1+2\epsilon}(1+x^2)^{-2\epsilon}\left(1+x^2\cot^2\frac{\theta_{ij}}{2}\right)^{-1}\nonumber\\
&
\hspace{1cm}
\int_{-1}^1dy\, (1+y)^{-1/2+\epsilon}(1-y)^{-1/2+\epsilon} \left(
1-\frac{2x\cot \frac{\theta_{ij}}{2}}{1+x^2\cot^2\frac{\theta_{ij}}{2}} y
\right)^{-1}\,.\nonumber
\end{align}
The integral over $y$ can be changed to range over $y\in[0,1]$:
\begin{align}
&\int_{-1}^1dy\, (1+y)^{-1/2+\epsilon}(1-y)^{-1/2+\epsilon} \left(
1-\frac{2x\cot \frac{\theta_{ij}}{2}}{1+x^2\cot^2\frac{\theta_{ij}}{2}} y
\right)^{-1}\\
&
=2^{2\epsilon}\frac{1+x^2\cot^2\frac{\theta_{ij}}{2}}{\left(1-x\cot \frac{\theta_{ij}}{2}\right)^2}\int_0^1dy\, y^{-1/2+\epsilon}(1-y)^{-1/2+\epsilon} \left(
1+\frac{4x\cot \frac{\theta_{ij}}{2}}{\left(1-x\cot \frac{\theta_{ij}}{2}\right)^2}\, y
\right)^{-1}\nonumber\\
&=\frac{\sqrt{\pi}\,\Gamma\left(
\frac{1}{2}+\epsilon
\right)}{\Gamma(1+\epsilon)}\frac{1+x^2\cot^2\frac{\theta_{ij}}{2}}{\left(1-x\cot \frac{\theta_{ij}}{2}\right)^2}\,_2F_1\left(
1,\frac{1}{2}+\epsilon,1+2\epsilon;-\frac{4x\cot \frac{\theta_{ij}}{2}}{\left(1-x\cot \frac{\theta_{ij}}{2}\right)^2}
\right)\,.
\nonumber
\end{align}
Here, $_2F_1(a,b,c;z)$ is the hypergeometric function.

Then, the remaining integral over $x$ is
\begin{align}
&2^{2\epsilon}\tan^{2\epsilon}\frac{\theta_{ij}}{2}\frac{\sqrt{\pi}\,\Gamma\left(
\frac{1}{2}+\epsilon
\right)}{\Gamma(1+\epsilon)}\int_0^\infty dx\, x^{-1+2\epsilon}|1-x|^{-1+2\epsilon}(1+x)^{-1-2\epsilon}\left(1+x^2\tan^2 \frac{\theta_{ij}}{2}\right)^{-2\epsilon}\nonumber\\
&
\hspace{6cm}
\times\,_2F_1\left(
2\epsilon,\frac{1}{2}+\epsilon,1+2\epsilon;\frac{4x}{\left(1+x\right)^2}
\right)\\
&
\simeq 2^{2\epsilon}\tan^{2\epsilon}\frac{\theta_{ij}}{2}\frac{\sqrt{\pi}\,\Gamma\left(
\frac{1}{2}+\epsilon
\right)}{\Gamma(1+\epsilon)}\left[
\int_0^1 dx\, x^{-1+2\epsilon}(1-x)^{-1+2\epsilon}(1+x)^{-1+2\epsilon}\left(1+x^2\tan^2 \frac{\theta_{ij}}{2}\right)^{-2\epsilon}
\right.\nonumber\\
&
\hspace{4cm}
\left.
+\int_1^\infty dx\, x^{-1-2\epsilon}(x-1)^{-1+2\epsilon}(1+x)^{-1+2\epsilon}\left(1+x^2\tan^2 \frac{\theta_{ij}}{2}\right)^{-2\epsilon}
\right]\,.
\nonumber
\end{align}
Here, the $\simeq$ symbol means that the expressions agree up through $\epsilon^0$ order.  To do this, we expanded the hypergeometric function as
\begin{align}
_2F_1\left(
2\epsilon,\frac{1}{2}+\epsilon,1+2\epsilon;\frac{4x}{\left(1+x\right)^2}
\right) = 1+4\epsilon \log\frac{2(1+x)}{1+x+|1-x|}+{\cal O}(\epsilon^2)\,.
\end{align}

This can be expressed as an integral on $x\in[0,1]$:
\begin{align}
&2^{2\epsilon}\tan^{2\epsilon}\frac{\theta_{ij}}{2}\frac{\sqrt{\pi}\,\Gamma\left(
\frac{1}{2}+\epsilon
\right)}{\Gamma(1+\epsilon)}\left[
\int_0^1 dx\, x^{-1+2\epsilon}(1-x)^{-1+2\epsilon}(1+x)^{-1+2\epsilon}\left(1+x^2\tan^2 \frac{\theta_{ij}}{2}\right)^{-2\epsilon}
\right.\nonumber\\
&
\hspace{4cm}
\left.
+\int_0^1 dx\, x^{-1+2\epsilon}(1-x)^{1+2\epsilon}(2-x)^{-1+2\epsilon}\left((1-x)^2+\tan^2 \frac{\theta_{ij}}{2}\right)^{-2\epsilon}
\right]\nonumber\\
&=2^{2\epsilon}\tan^{2\epsilon}\frac{\theta_{ij}}{2}\frac{\sqrt{\pi}\,\Gamma\left(
\frac{1}{2}+\epsilon
\right)}{\Gamma(1+\epsilon)}\left[
\int_0^1 dx\, x^{-1+2\epsilon}(1-x)^{-1+2\epsilon}(1+x)^{-1+2\epsilon}\left(1+x^2\tan^2 \frac{\theta_{ij}}{2}\right)^{-2\epsilon}
\right.\nonumber\\
&
\hspace{8cm}
\left.
+\frac{1}{4\epsilon}-\frac{1}{2}\log\left(
1+\tan^2 \frac{\theta_{ij}}{2}
\right)
\right]
\,.
\nonumber
\end{align}

The first integral can be broken into two regions, corresponding to the two divergences at $x=0$ and $x=1$:
\begin{align}
&\int_0^1 dx\, x^{-1+2\epsilon}(1-x)^{-1+2\epsilon}(1+x)^{-1+2\epsilon}\left(1+x^2\tan^2 \frac{\theta_{ij}}{2}\right)^{-2\epsilon}\\
&
\hspace{1cm}
=2^{2-2\epsilon}\int_0^1 dx\, x^{-1+2\epsilon}(2-x)^{-1+2\epsilon}(2+x)^{-1+2\epsilon}\left(4+x^2\tan^2 \frac{\theta_{ij}}{2}\right)^{-2\epsilon}\nonumber\\
&\hspace{2cm}
+2^{2-6\epsilon}\int_0^1 dx\, x^{-1+2\epsilon}(2-x)^{-1+2\epsilon}(4-x)^{-1+2\epsilon}\left(1+\left(1-\frac{x}{2}\right)^2\tan^2 \frac{\theta_{ij}}{2}\right)^{-2\epsilon}\nonumber\\
&
\hspace{1cm}
=\frac{3}{4}\frac{1}{\epsilon} -\frac{1}{2}\log\left(
1+\tan^2 \frac{\theta_{ij}}{2}
\right)
\,.
\nonumber
\end{align}

Then, putting it together, we find
\begin{align}\label{eq:softfunccalc}
&2^{2\epsilon}\tan^{2\epsilon}\frac{\theta_{ij}}{2}\frac{\sqrt{\pi}\,\Gamma\left(
\frac{1}{2}+\epsilon
\right)}{\Gamma(1+\epsilon)}\left[
\frac{3}{4}\frac{1}{\epsilon} -\frac{1}{2}\log\left(
1+\tan^2 \frac{\theta_{ij}}{2}
\right)+\frac{1}{4\epsilon}-\frac{1}{2}\log\left(
1+\tan^2 \frac{\theta_{ij}}{2}
\right)
\right]\\
&
\hspace{1cm}
=\pi \left[
\frac{1}{\epsilon} +\log(1-\cos\theta_{ij})-\log 2
\right]
\nonumber
\end{align}

Then, we find in total (in $\overline{\text{MS}}$)
\begin{align}\label{eq:softfunctot}
S_{ij}(D)&=-2\frac{\alpha_s}{\pi}{\bf T}_i\cdot {\bf T}_j \left(
\frac{81 C^2\mu^2}{Q^2}
\right)^\epsilon \frac{1}{D^{1+2\epsilon}}\left(
\frac{1}{\epsilon}+\log \frac{1-x_k}{4 x_i x_j}
\right)\,.
\end{align}
Laplace-transforming in the $D$-parameter, we have
\begin{align}
S_{ij}(\tilde D)&=\frac{\alpha_s}{\pi}{\bf T}_i\cdot {\bf T}_j  \left(
\frac{1}{\epsilon^2}+\frac{1}{\epsilon}\log\frac{81 C^2 \tilde D^2\mu^2}{Q^2}+\frac{1}{\epsilon}\log \frac{1-x_k}{4 x_i x_j}
\right)\,.
\end{align}
The anomalous dimension is therefore
\begin{equation}
\gamma_{S_{ij}} = 2\frac{\alpha_s}{\pi}{\bf T}_i\cdot {\bf T}_j \log\frac{81 C^2 \tilde D^2\mu^2}{4Q^2}+2\frac{\alpha_s}{\pi}{\bf T}_i\cdot {\bf T}_j \log \frac{1-x_k}{x_i x_j}\,.
\end{equation}

The color matrix products for $e^+e^-\to q\bar q g$ are:
\begin{equation}
{\bf T}_q\cdot {\bf T}_g = -\frac{C_A}{2}\,,\qquad
{\bf T}_{\bar q}\cdot {\bf T}_g = -\frac{C_A}{2}\,,\qquad
{\bf T}_q\cdot {\bf T}_{\bar q} = \frac{C_A}{2}-C_F\,.
\end{equation}
We can then sum over the colors to find:
\begin{align}
\gamma_S &= -\frac{\alpha_s}{\pi}(C_A+2C_F)\log\frac{81 C^2 \tilde D^2\mu^2}{4Q^2}+2\frac{\alpha_s C_F}{\pi}\log x_1 x_2+2\frac{\alpha_s C_A}{\pi}\log(2-x_1-x_2)\\
&
\hspace{2cm}
-\frac{\alpha_s C_A}{\pi}\log(1-x_1)-\frac{\alpha_s C_A}{\pi}\log(1-x_2)+\frac{\alpha_s}{\pi}(C_A-2C_F)\log(x_1+x_2-1)\,.
\nonumber
\end{align}

This can be expressed in terms of the $C$-parameter in the suggestive form:
\begin{align}
\gamma_S &= -\frac{\alpha_s}{\pi}(C_A+2C_F)\log\frac{81 C^2 \tilde D^2\mu^2}{4Q^2}+\frac{\alpha_s}{\pi}(2C_F-C_A)\log \frac{x_1 x_2}{x_1+x_2-1}\\
&
\hspace{2cm}
-\frac{\alpha_s C_A}{\pi}\log\frac{C}{6}\frac{1}{(x_1+x_2-1)(2-x_1-x_2)}\,.
\nonumber
\end{align}

\subsection{Summary of Anomalous Dimensions}\label{app:anomsumm}

Here, we summarize the anomalous dimensions for this region.  We have:
\begin{align}
\gamma_H &= -2\frac{\alpha_sC_F}{\pi}\log\frac{\mu^2}{Q^2}-3\frac{\alpha_s C_F}{\pi}-\frac{\alpha_sC_A}{\pi}\log\frac{6\mu^2}{(2-x_1-x_2)CQ^2}-\frac{\alpha_s}{\pi}\frac{11C_A-2 n_f}{6}\\
&
\hspace{1cm}
-\frac{\alpha_sC_A}{\pi}\log\frac{x_1+x_2-1}{x_1x_2}-\frac{\alpha_s}{\pi}(C_A-2C_F)\log(x_1+x_2-1)\,,\nonumber\\
\gamma_{q} &= 2\frac{\alpha_sC_F}{\pi} \log \frac{9C \tilde D\mu^2}{2x_1 Q^2}+\frac{3}{2}\frac{\alpha_sC_F}{\pi}\,,\nonumber \\
\gamma_{\bar q} &= 2\frac{\alpha_sC_F}{\pi} \log \frac{9C \tilde D\mu^2}{2x_2 Q^2}+\frac{3}{2}\frac{\alpha_sC_F}{\pi}\,,\nonumber \\
\gamma_{g} &= 2\frac{\alpha_s C_A}{\pi} \log\frac{9 C \tilde D\mu^2}{2(2-x_1-x_2) Q^2}+\frac{\alpha_s}{\pi}\frac{11C_A-2n_f}{6}\,,\nonumber\\
\gamma_S &= -\frac{\alpha_s}{\pi}(C_A+2C_F)\log\frac{81 C^2 \tilde D^2\mu^2}{4Q^2}+\frac{\alpha_s}{\pi}(2C_F-C_A)\log \frac{x_1 x_2}{x_1+x_2-1}\\
&
\hspace{2cm}
-\frac{\alpha_s C_A}{\pi}\log\frac{C}{6}\frac{1}{(x_1+x_2-1)(2-x_1-x_2)}\,. \nonumber
\end{align}
As required, the sum of these anomalous dimensions is 0.

\section{Region 2 Calculations}\label{app:reg2app}

There are two configurations that contribute to region 2, when $D\ll C^2 \ll 1$.  The emissions that set the value of the $C$-parameter can be hard and collinear or they can be soft and wide-angle.  We will consider each of these in turn.

\subsection{Collinear Region}\label{app:collreg2anom}

The factorization theorem for the collinear contributions is
\begin{equation}
\frac{1}{\sigma}\frac{d^3\sigma^\text{coll}}{dC\,dD\, dx_2} = H(Q^2) H_{1\to 2}(x_2,C)J_1(D)  J_2(x_2,D)J_3(1-x_2,D) S_{12}(C,D)C_s(C,D,x_2)\,.
\end{equation}

\subsubsection{Hard Function, $H(Q^2)$}

The one-loop hard function for $e^+e^-\to q\bar q$ production is \cite{Bauer:2003di,Manohar:2003vb,Ellis:2010rwa,Bauer:2011uc}
\begin{equation}\label{eq:hard12def}
H(Q^2) = 1 + \frac{\alpha_s\,C_F}{2\pi} \left(-\log^2\frac{\mu^2}{Q^2} - 3\log \frac{\mu^2}{Q^2} - 8+\frac{7}{6}\pi^2\right)\,.
\end{equation}
Its anomalous dimension is thus
\begin{equation}
\gamma_H = -2\frac{\alpha_s C_F}{\pi}\log\frac{\mu^2}{Q^2}-3\frac{\alpha_s C_F}{\pi}\,.
\end{equation}

\subsubsection{Hard Splitting Function, $H_{1\to 2}(x_2,C)$}

The leading-order hard splitting function for collinear gluon emission off of the anti-quark is
\begin{align}
H_{1\to 2}(x_2,C)=\frac{\alpha_s C_F}{2\pi}\frac{1}{C}\frac{1+x_2^2}{1-x_2}\,, 
\end{align}
where $x_2$ is the energy fraction of the anti-quark in the splitting.  To write this, we have used the expression for the $C$-parameter in this region as
\begin{equation}
C=6(1-x_1)\,.
\end{equation}
Its one-loop anomalous dimension is \cite{Bauer:2011uc,Ellis:1980wv,Kosower:1999rx}
\begin{align}
\gamma_{H_{1\to 2}} = -\frac{\alpha_s C_A}{\pi}\log\frac{6\mu^2}{(1-x_2)C Q^2}+\frac{\alpha_s}{\pi}(2C_F-C_A)\log x_2 - \frac{\alpha_s}{\pi}\frac{11 C_A-2n_f}{6}\,.
\end{align}

\subsubsection{Wide-angle Soft Function, $S_{12}(D,C)$}

We can use the results from the end of \Sec{sec:sfreg1} to evaluate the anomalous dimensions of the wide-angle soft function, $S_{12}(C,D)$.  The product of color factors is
\begin{equation}
{\bf T}_1\cdot {\bf T}_2 = -C_F
\end{equation}
and the angle between the two jets in the event is $1-\cos\theta_{12} = 2$.  Therefore, its anomalous dimension in Laplace space is
\begin{equation}\label{eq:wasoftreg2}
\gamma_{S_{12}} = -2\frac{\alpha_s C_F}{\pi} \log\frac{81 C^2 \tilde D^2\mu^2}{4Q^2}\,.
\end{equation}

\subsubsection{Collinear-Soft Function, $C_{s}(C,D,x_2)$}

The anomalous dimension of the collinear-soft function $C_{s}(C,D,x_2)$ can also be found from the results at the end of \Sec{sec:sfreg1}.  We take the $x_1\to 1$ limit of the soft function anomalous dimension in region 1:
\begin{align}
\gamma_{S}&\xrightarrow[x_1\to1]{} -\frac{\alpha_s}{\pi}(C_A+2C_F)\log\frac{81 C^2 \tilde D^2\mu^2}{4Q^2}-\frac{\alpha_s C_A}{\pi}\log\frac{C}{6}\frac{1}{x_2(1-x_2)}\,.
\end{align}
Then, subtracting the wide-angle soft function's anomalous dimension of \Eq{eq:wasoftreg2}, we find
\begin{equation}
\gamma_{C_s} = -\frac{\alpha_sC_A}{\pi}\log\frac{81 C^3 \tilde D^2\mu^2}{4Q^2}+\frac{\alpha_s C_A}{\pi}\log\left(6x_2(1-x_2)\right)\,.
\end{equation}

\subsubsection{Summary of Anomalous Dimensions}

Here, we summarize the anomalous dimensions for this region.  We have:
\begin{align}
\gamma_H &= -2\frac{\alpha_s C_F}{\pi}\log\frac{\mu^2}{Q^2}-3\frac{\alpha_s C_F}{\pi}\,,\\
\gamma_{H_{1\to 2}} &= -\frac{\alpha_s C_A}{\pi}\log\frac{6\mu^2}{(1-x_2)C Q^2}+\frac{\alpha_s}{\pi}(2C_F-C_A)\log x_2 - \frac{\alpha_s}{\pi}\frac{11 C_A-2n_f}{6}\,,\nonumber\\
\gamma_{q} &= 2\frac{\alpha_sC_F}{\pi} \log \frac{9C \tilde D\mu^2}{2Q^2}+\frac{3}{2}\frac{\alpha_sC_F}{\pi}\,,\nonumber \\
\gamma_{\bar q} &= 2\frac{\alpha_sC_F}{\pi} \log \frac{9C \tilde D\mu^2}{2x_2 Q^2}+\frac{3}{2}\frac{\alpha_sC_F}{\pi}\,,\nonumber \\
\gamma_{g} &= 2\frac{\alpha_s C_A}{\pi} \log\frac{9 C \tilde D\mu^2}{2(1-x_2) Q^2}+\frac{\alpha_s}{\pi}\frac{11C_A-2n_f}{6}\,,\nonumber\\
\gamma_{S_{12}} &= -2\frac{\alpha_s C_F}{\pi} \log\frac{81 C^2 \tilde D^2\mu^2}{4Q^2}\,, \nonumber\\
\gamma_{C_s} &= -\frac{\alpha_sC_A}{\pi}\log\frac{81 C^3 \tilde D^2\mu^2}{4Q^2}+\frac{\alpha_s C_A}{\pi}\log\left(6x_2(1-x_2)\right)\,.\nonumber
\end{align}
As required, the sum of these anomalous dimensions is 0.

\subsection{Soft Region}\label{app:softreg2anom}

The factorization theorem for the soft contribution is
\begin{equation}
\frac{1}{\sigma}\frac{d^3\sigma^\text{soft}}{dC\, dD\, dx_3} = H(Q^2) H_s(x_3,C)J_1(D)J_2(D)  J_3(x_3,D) S_{123}(x_3,C,D)\,.
\end{equation}
The hard function, $H(Q^2)$, is defined above in \Eq{eq:hard12def}.  The leading-order hard splitting function for soft gluon emission off of a quark and anti-quark dipole $H_s(x_3,C)$ can be found from changing variables in the soft limit of the $e^+e^-\to q\bar q g$ matrix element.  We have
\begin{align}
&\int_0^1dx_1 \int_0^1 dx_2\,\frac{\alpha_sC_F}{\pi}\frac{1}{(1-x_1)(1-x_2)}\delta\left(
C-6\frac{(1-x_1)(1-x_2)}{2-x_1-x_2}
\right)\delta(x_3-(2-x_1-x_2))\nonumber \\
&
\hspace{1cm}
=2\frac{\alpha_s C_F}{\pi}\frac{1}{x_3C\sqrt{1-\frac{2C}{3x_3}}}\,.
\end{align}
That is, the hard splitting function is
\begin{equation}
H_s(x_3,C)=2\frac{\alpha_s C_F}{\pi}\frac{1}{x_3C\sqrt{1-\frac{2C}{3x_3}}}\,,
\end{equation}
where $x_3=2-x_1-x_2$ is the energy fraction of the gluon.  The anomalous dimension is \cite{Catani:2000pi,Larkoski:2015zka}
\begin{align}
\gamma_{H_s} &= -\frac{\alpha_sC_A}{\pi}\log\frac{6\mu^2}{x_3 CQ^2}-\frac{\alpha_s}{\pi}\frac{11C_A-2 n_f}{6}\,.
\end{align}
All other functions in this region are just the $x_3\to 0$ limits of the corresponding functions from region 1.

\section{Region 3 Calculations}\label{app:reg3}

\subsection{$C$-parameter at NLL accuracy}

The $C$-parameter was first resummed to NLL accuracy in \Ref{Catani:1998sf}, by relating its value in the limit as $C\to 0$ to thrust at NLL from \Ref{Catani:1991kz}.  The cumulative distribution at NLL accuracy is
\begin{equation}
\Sigma(C) = \frac{e^{-R(C,\mu)-\gamma_E S(C)}}{\Gamma\left(1+S(C)\right)}\,.
\end{equation}
Here, the function $R(C,\mu)$ is the radiator, which to NLL is
\begin{align}
R(C,\mu) &= \frac{16\pi C_F}{\alpha_s \beta_0^2}\left[
\left(
1-\lambda(C)
\right)\log\left(
1-\lambda(C)
\right)
-2\left(
1-\frac{\lambda(C)}{2}
\right)\log\left(
1-\frac{\lambda(C)}{2}
\right)
\right]\\
&
\hspace{-1cm}
-\frac{8C_F}{\beta_0^2}\left[
\beta_0\log\frac{Q}{\mu}-\left(
\left(
\frac{67}{18}-\frac{\pi^2}{6}
\right)C_A-\frac{10}{9}n_f T_R
\right)
\right]\log\frac{\left(1-\frac{\lambda(C)}{2}\right)^2}{1-\lambda(C)}+\frac{6C_F}{\beta_0}\log\left(
1-\frac{\lambda(C)}{2}
\right)
\nonumber\\
&
\hspace{-1cm}
+\frac{4C_F\beta_1}{\beta_0^3}\left[
\log\left(
1-\frac{\lambda(C)}{2}
\right)-\log\left(
1-\lambda(C)
\right)+\frac{1}{2}\log^2\left(
1-\lambda(C)
\right)-\log^2\left(
1-\frac{\lambda(C)}{2}
\right)
\right]\,.
\nonumber
\end{align}
$\Gamma(x)$ is the Euler Gamma function, and $\gamma_E=0.5772\dotsc$ is the Euler-Mascheroni constant.  The function $\lambda(C)$ is
\begin{equation}
\lambda(C) = \frac{\alpha_s}{2\pi}\beta_0\log\frac{6}{C}\,.
\end{equation}
and the function $S(C)$ is
\begin{align}
S(C) = \frac{8C_F}{\beta_0}
\log\left(
\frac{1-\frac{\lambda(C)}{2}}{1-\lambda(C)}
\right)
\end{align}
$\beta_0$ and $\beta_1$ are the one- and two-loop $\beta$-function coefficients which are
\begin{align}
\beta_0 &=\frac{11}{3}C_A -\frac{4}{3}T_R n_f\,,\\
\beta_1& =\frac{34}{3}C_A^2-4T_R n_f\left(
C_F+\frac{5}{3}C_A
\right)\,.
\end{align}
The coupling $\alpha_s\equiv \alpha_s(\mu)$ is evaluated at the scale $\mu$, which is of the order of the center-of-mass collision energy, $Q$.

The $C$-parameter has been resummed to N$^3$LL accuracy by relating it to thrust by factorizing the cross section with soft-collinear effective theory \cite{Hoang:2015hka}.  While such a factorization is useful for resumming the $D$-parameter in regions 1 and 2, for compactness here, we prefer to use the form from the earlier literature.  The differential cross section of the $C$-parameter is just the derivative of this expression:
\begin{align}
\frac{1}{\sigma}\frac{d\sigma^\text{NLL}}{dC} = \frac{\partial}{\partial C}\Sigma(C)\,.
\end{align}

For calculating the conditional cross section of the $D$-parameter given $C$, we also need the cross section expanded to lowest order in $\alpha_s$.  Here, we choose to include resummation of collinear logarithms through one-loop running of the coupling as well, as this will result in successfully resumming more logarithms of $D$.  This can be calculated by simply differentiating the radiator with respect to $C$ and keeping those terms that correspond to one-loop running.  That is, the lowest-order differential cross section of the $C$-parameter including the effects of one-loop running of $\alpha_s$ is
\begin{align}
\frac{1}{\sigma}\frac{d\sigma^{\text{1-loop }\alpha_s}}{dC} &= -\frac{\partial R(C,\mu)}{\partial C}=\frac{8C_F}{\beta_0}\frac{1}{C}\log\frac{1-\frac{\lambda(C)}{2}}{1-\lambda(C)}-\frac{3}{2}\frac{\alpha_s C_F}{2\pi}\frac{1}{C}\frac{1}{1-\frac{\lambda(C)}{2}}\,.
\end{align}
Expanding this to lowest order in $\alpha_s$ produces the result in \Eq{eq:singas}.

\subsection{Double Differential Cross Section for $C$- and $D$-parameters}\label{sec:cdreg3appcalc}

To calculate the $C$ and $D$ parameters in this region of phase space we have
\begin{align}
\frac{1}{\sigma_0}\frac{d^2\sigma}{dC\, dD} = \int[d^dk_1]_+[d^dk_2]_+\, |{\cal M}(k_1,k_2)|^2\,\delta(C-\hat C)\delta(D-\hat D)\,.
\end{align}
It is most convenient to express the phase space integrals in terms of energy $E$, polar angle $\theta$, and azimuthal angle $\phi$ for both emissions.  In the soft and collinear limit, the phase space integral is
\begin{equation}
\int [d^dk]_+ = \frac{1}{8\pi^3}\int_0^{Q/2} dE\, E\int_0^1 d\theta\, \theta\int_0^\pi d\phi\,.
\end{equation}
The maximum value of the energy in the soft and collinear limit is arbitrary, but we will find that the choice of $E_{\max}=Q/2$ is convenient.  Only the relative angle between the particles $1$ and $2$ is important, so we can integrate over one the particles' $\phi$ angles.  Doing this, we find
\begin{equation}
\int [d^dk]_+ = \frac{1}{8\pi^2}\int_0^{Q/2} dE\,E \int_0^1 d\theta\, \theta\,.
\end{equation}
Then, the integrals we need to do are
\begin{align}
\frac{1}{\sigma_0}\frac{d^2\sigma}{dC\, dD} &= \frac{1}{64\pi^5} \int_0^{Q/2} dE_1\,E_1 \int_0^1 d\theta_1\, \theta_1\\
&
\hspace{1cm}
\times \int_0^{Q/2} dE_2\,E_2 \int_0^1 d\theta_2\, \theta_2\int_0^\pi d\phi\, |{\cal M}(k_1,k_2)|^2\,\delta(C-\hat C)\delta(D-\hat D)\,.
\nonumber
\end{align}

Now, we would like to pull out all of the scales from the $\delta$-functions, so the integrands are purely ${\cal O}(1)$.  First, energy $Q$ dependence is trivial:
\begin{align}
\frac{1}{\sigma_0}\frac{d^2\sigma}{dC\, dD} &= \frac{1}{64\pi^3} \int_0^1 dz_1\,z_1 \int_0^1 d\theta_1\, \theta_1\\
&
\hspace{1cm}
\times \int_0^1 dz_2\,z_2 \int_0^1 d\theta_2\, \theta_2\int_0^\pi d\phi\, \alpha_s(k_{\perp 1})\alpha_s(k_{\perp 2})|{\cal M}(k_1,k_2)|^2\,\delta(C-\hat C)\delta(D-\hat D)\,,
\nonumber
\end{align}
where $z_1$ and $z_2$ are now energy fractions with respect to $Q/2$.  In doing this, we have explicitly written the $\alpha_s$ dependence, multiplying by $(4\pi \alpha_s)^2$.  Also, to resum collinear logarithms, we evaluate the coupling at the transverse momenta of particles 1 and 2.  To NLL accuracy, we only need one-loop running of the coupling, which is
\begin{equation}
\alpha_s(\mu) = \frac{\alpha_s(Q)}{1+\frac{\alpha_s(Q)}{2\pi}\beta_0\log\frac{\mu}{Q}}\,.
\end{equation}
Here, $Q$ is a reference scale, which we will take to be the center-of-mass collision energy.  $\beta_0$ is the one-loop coefficient of the QCD $\beta$-function
\begin{equation}
\beta_0 =\frac{11}{3}C_A -\frac{4}{3}T_R n_f\,.
\end{equation}
Because region 3 corresponds to $D\sim C^2$ or that $k_{\perp1}\sim k_{\perp 2}$, to NLL accuracy, we can just evaluate the couplings at the same scale.  So, we have
\begin{align}
\frac{1}{\sigma_0}\frac{d^2\sigma}{dC\, dD} &= \frac{1}{64\pi^3} \int_0^1 dz_1\,z_1 \int_0^1 d\theta_1\, \theta_1\\
&
\hspace{1cm}
\times \int_0^1 dz_2\,z_2 \int_0^1 d\theta_2\, \theta_2\int_0^\pi d\phi\, \alpha_s(k_{\perp 1})^2|{\cal M}(k_1,k_2)|^2\,\delta(C-\hat C)\delta(D-\hat D)\,,
\nonumber
\end{align}
where $k_{\perp1} = z_1\theta_1 \frac{Q}{2}$.

With this rescaling, the $C$ and $D$ parameters are
\begin{align}
C&=\frac{3}{2}\left(z_1\theta_1^2+z_2\theta_2^2\right)\,,\\
D&=\frac{27}{4}z_1\theta_1^2z_2 \theta_2^2\sin^2\phi\,.\nonumber
\end{align}
So, the cross section with the explicit $\delta$-functions is
\begin{align}
\frac{1}{\sigma_0}\frac{d^2\sigma}{dC\, dD} &= \frac{1}{64\pi^3} \int_0^1 dz_1\,z_1 \int_0^1 d\theta_1\, \theta_1\int_0^1 dz_2\,z_2 \int_0^1 d\theta_2\, \theta_2\int_0^\pi d\phi\, \alpha_s(z_1\theta_1Q)^2|{\cal M}(k_1,k_2)|^2\\
&
\hspace{1cm}
\times \delta\left(C-\frac{3}{2}\left(z_1\theta_1^2+z_2\theta_2^2\right)\right)\delta\left(D-\frac{27}{4}z_1\theta_1^2z_2 \theta_2^2\sin^2\phi\right)\nonumber\,.
\end{align}

Now, we will make the change of variables
\begin{align}\label{eq:kperpchange}
&z_2 = uz_1\,, &\theta_2 = v\theta_1\,.
\end{align}
The cross section then becomes
\begin{align}
\frac{1}{\sigma_0}\frac{d^2\sigma}{dC\, dD} &= \frac{1}{64\pi^3} \int_0^\infty \frac{dz_1}{z_1} \int_0^\infty \frac{d\theta_1}{\theta_1}\int_0^\infty du\,u \int_0^\infty dv\, v\int_0^\pi d\phi\, \alpha_s(z_1\theta_1Q)^2 |{\cal M}(k_1,k_2)|^2\\
&
\hspace{-1cm}
\times \delta\left(C-\frac{3}{2}z_1\theta_1^2\left(1+uv^2\right)\right)\delta\left(D-\frac{27}{4}z_1^2\theta_1^4uv^2\sin^2\phi\right)\Theta(1-uz_1)\Theta(1-v\theta_1)\Theta(1-z_1)\Theta(1-\theta_1)\nonumber\,.
\end{align}
To write this expression, we note that in the soft and collinear limit, the matrix element $|{\cal M}(k_1,k_2)|^2\propto z_1^{-4}\theta_1^{-4}$, which we have pulled out and made explicit.  Now, using the $C$-parameter $\delta$-function, we can do the integral over $z_1$.  We find
\begin{align}
\frac{1}{\sigma_0}\frac{d^2\sigma}{dC\, dD} &= \frac{1}{64\pi^3}\frac{1}{C^3} \int_0^\infty \frac{d\theta_1}{\theta_1}\int_0^\infty du\,u \int_0^\infty dv\, v\int_0^\pi d\phi\, \alpha_s\left(\frac{2C}{3\theta_1^2(1+uv^2)}\theta_1Q\right)^2 |{\cal M}(k_1,k_2)|^2\\
&
\hspace{-1cm}
\times \delta\left(\frac{D}{C^2}-\frac{3uv^2\sin^2\phi}{(1+uv^2)^2}\right)\Theta\left(\theta_1^2-\frac{2uC}{3(1+uv^2)}\right)\Theta\left(\frac{1}{v}-\theta_1\right)\Theta\left(\theta_1^2-\frac{2C}{3(1+uv^2)}\right)\Theta(1-\theta_1)\nonumber\,.
\end{align}
The integral over $\theta_1$ can then be done as the only non-trivial dependence is in the running coupling.  We can simplify the expression for the running coupling as follows.  By selection of this region, $uv^2\sim 1$ and so to NLL accuracy, we can just ignore the multiplicative factors in the running coupling scale.  That is, we can simplify this expression to
\begin{align}
\frac{1}{\sigma_0}\frac{d^2\sigma}{dC\, dD} &= \frac{1}{64\pi^3}\frac{1}{C^3} \int_0^\infty \frac{d\theta_1}{\theta_1}\int_0^\infty du\,u \int_0^\infty dv\, v\int_0^\pi d\phi\, \alpha_s\left(\frac{CQ}{\theta_1}\right)^2 |{\cal M}(k_1,k_2)|^2\\
&
\hspace{-1cm}
\times \delta\left(\frac{D}{C^2}-\frac{3uv^2\sin^2\phi}{(1+uv^2)^2}\right)\Theta\left(\theta_1^2-\frac{2uC}{3(1+uv^2)}\right)\Theta\left(\frac{1}{v}-\theta_1\right)\Theta\left(\theta_1^2-\frac{2C}{3(1+uv^2)}\right)\Theta(1-\theta_1)\nonumber\,.
\end{align}

To continue, the phase space constraints imposed by the $\Theta$-functions can be manipulated into
\begin{align}
\Theta\left(\theta_1^2-\frac{2uC}{3(1+uv^2)}\right)\Theta\left(\frac{1}{v}-\theta_1\right)\Theta\left(\theta_1^2-\frac{2C}{3(1+uv^2)}\right)\Theta(1-\theta_1)&=\\
&
\hspace{-6cm}
\Theta\left(\theta_1^2-\frac{2C}{3(1+uv^2)}\right)\Theta(1-\theta_1)\Theta(1-v)\Theta(1-u)\nonumber\\
&
\hspace{-6cm}
+\Theta\left(\theta_1^2-\frac{2uC}{3(1+uv^2)}\right)\Theta(1-\theta_1)\Theta(1-v)\Theta(u-1)\nonumber\\
&
\hspace{-6cm}
+\Theta\left(\theta_1^2-\frac{2C}{3(1+uv^2)}\right)\Theta\left(\frac{1}{v}-\theta_1\right)\Theta(v-1)\Theta(1-u)\nonumber\\
&
\hspace{-6cm}
+\Theta\left(\theta_1^2-\frac{2uC}{3(1+uv^2)}\right)\Theta\left(\frac{1}{v}-\theta_1\right)\Theta(v-1)\Theta(u-1)\nonumber\,.
\end{align}
Then, to NLL accuracy, by integrating over $\theta_1$, the cross section becomes
\begin{align}
\frac{1}{\sigma_0}\frac{d^2\sigma}{dC\, dD} &= \frac{1}{128\pi^3}\frac{1}{C^3}\int_0^\infty du\,u \int_0^\infty dv\, v\int_0^\pi d\phi\, \alpha_s(CQ)\alpha_s(\sqrt{C}Q) |{\cal M}(k_1,k_2)|^2\\
&
\hspace{-1cm}
\times \delta\left(\frac{D}{C^2}-\frac{3uv^2\sin^2\phi}{(1+uv^2)^2}\right)\left[
\Theta(1-v)\Theta(1-u)\Theta(3(1+uv^2)-2C)\log\frac{3(1+uv^2)}{2C}\right.\nonumber\\
&
\hspace{2cm}
\left.
+\Theta(1-v)\Theta(u-1)\Theta(3(1+uv^2)-2uC)\log\frac{3(1+uv^2)}{2uC}\right.\nonumber\\
&
\hspace{2cm}
\left.
+\Theta(v-1)\Theta(1-u)\Theta(3(1+uv^2)-2v^2C)\log\frac{3(1+uv^2)}{2v^2C}\right.\nonumber\\
&
\hspace{2cm}
\left.
+\Theta(v-1)\Theta(u-1)\Theta(3(1+uv^2)-2uv^2C)\log\frac{3(1+uv^2)}{2uv^2C}\right]\nonumber\,.
\end{align}
To evaluate this expression, we have used the behavior of the integral over the running coupling
\begin{align}
\int_{C^{1/2}}^1\frac{d\theta_1}{\theta_1}\alpha_s\left(\frac{CQ}{\theta_1}\right)^2 = -\alpha_s(CQ)\alpha_s\left(\sqrt{C}Q\right)\frac{\log C}{2}\,.
\end{align}
In what follows, we use the shorthand notation
\begin{equation}
\alpha_s^2\equiv \alpha_s(CQ)\alpha_s\left(\sqrt{C}Q\right)\,.
\end{equation}

To evaluate the remaining $\delta$-function integral, we will integrate over $\phi$.  To do this, it is convenient to change variables to
\begin{equation}
x=\cos\phi\,,
\end{equation}
so that the cross section becomes
\begin{align}
\frac{1}{\sigma_0}\frac{d^2\sigma}{dC\, dD} &= \frac{\alpha_s^2}{128\pi^3}\frac{1}{C^3}\int_0^\infty du\,u \int_0^\infty dv\, v\int_{-1}^1 dx\,(1-x^2)^{-1/2}  |{\cal M}(k_1,k_2)|^2\\
&
\hspace{-1cm}
\times \delta\left(\frac{D}{C^2}-\frac{3uv^2(1-x^2)}{(1+uv^2)^2}\right)\left[
\Theta(1-v)\Theta(1-u)\Theta(3(1+uv^2)-2C)\log\frac{3(1+uv^2)}{2C}\right.\nonumber\\
&
\hspace{2cm}
\left.
+\Theta(1-v)\Theta(u-1)\Theta(3(1+uv^2)-2uC)\log\frac{3(1+uv^2)}{2uC}\right.\nonumber\\
&
\hspace{2cm}
\left.
+\Theta(v-1)\Theta(1-u)\Theta(3(1+uv^2)-2v^2C)\log\frac{3(1+uv^2)}{2v^2C}\right.\nonumber\\
&
\hspace{2cm}
\left.
+\Theta(v-1)\Theta(u-1)\Theta(3(1+uv^2)-2uv^2C)\log\frac{3(1+uv^2)}{2uv^2C}\right]\nonumber\,.
\end{align}
Doing the integral over $x$, we then find
\begin{align}
\frac{1}{\sigma_0}\frac{d^2\sigma}{dC\, dD} &= \frac{\alpha_s^2}{256\pi^3}\frac{1}{C D}\int_0^\infty du\,u \int_0^\infty dv\, v\,  \left(|{\cal M}(x=A)|^2+|{\cal M}(x=-A)|^2\right)\\
&
\hspace{-1.5cm}
\times \Theta\left(\frac{C^2}{D}-\frac{(1+uv^2)^2}{3uv^2}\right)\left(
\frac{C^2}{D}\frac{3uv^2}{(1+uv^2)^2}-1
\right)^{-1/2}\left[
\Theta(1-v)\Theta(1-u)\Theta(3(1+uv^2)-2C)\log\frac{3(1+uv^2)}{2C}\right.\nonumber\\
&
\hspace{2cm}
\left.
+\Theta(1-v)\Theta(u-1)\Theta(3(1+uv^2)-2uC)\log\frac{3(1+uv^2)}{2uC}\right.\nonumber\\
&
\hspace{2cm}
\left.
+\Theta(v-1)\Theta(1-u)\Theta(3(1+uv^2)-2v^2C)\log\frac{3(1+uv^2)}{2v^2C}\right.\nonumber\\
&
\hspace{2cm}
\left.
+\Theta(v-1)\Theta(u-1)\Theta(3(1+uv^2)-2uv^2C)\log\frac{3(1+uv^2)}{2uv^2C}\right]\nonumber\,.
\end{align}
Here, $A$ is the solution of the $\delta$-function:
\begin{equation}
A=\left(
1-\frac{D}{C^2}\frac{(1+uv^2)^2}{3uv^2}
\right)^{1/2}
\end{equation}

It is useful to change variables to $D$ and $y$, where
\begin{equation}
y=\frac{D}{C^2}\,.
\end{equation}
Then, $C=D^{1/2}y^{-1/2}$ and the Jacobian from this change of variables is
\begin{equation}
J=\frac{1}{2}D^{1/2}y^{-3/2}\,.
\end{equation}
With this change, the cross section becomes
\begin{align}
\frac{1}{\sigma_0}\frac{d^2\sigma}{dD\, dy} &= \frac{\alpha_s^2}{1028\pi^3}\frac{1}{y D}\int_0^\infty du\,u \int_0^\infty dv\, v\,  \left(|{\cal M}(x=A)|^2+|{\cal M}(x=-A)|^2\right)\\
&
\hspace{4cm}
\times \Theta\left(\frac{1}{y}-\frac{(1+uv^2)^2}{3uv^2}\right)\left(
\frac{1}{y}\frac{3uv^2}{(1+uv^2)^2}-1
\right)^{-1/2}\nonumber\\
&
\hspace{2cm}\times\left[
\Theta(1-v)\Theta(1-u)\Theta\left(3(1+uv^2)-2\sqrt{\frac{D}{y}}\right)\log\frac{9y(1+uv^2)^2}{4D}\right.\nonumber\\
&
\hspace{2cm}
\left.
+\Theta(1-v)\Theta(u-1)\Theta\left(3(1+uv^2)-2u\sqrt{\frac{D}{y}}\right)\log\frac{9y(1+uv^2)^2}{4u^2D}\right.\nonumber\\
&
\hspace{2cm}
\left.
+\Theta(v-1)\Theta(1-u)\Theta\left(3(1+uv^2)-2v^2\sqrt{\frac{D}{y}}\right)\log\frac{9y(1+uv^2)^2}{4v^4D}\right.\nonumber\\
&
\hspace{2cm}
\left.
+\Theta(v-1)\Theta(u-1)\Theta\left(3(1+uv^2)-2uv^2\sqrt{\frac{D}{y}}\right)\log\frac{9y(1+uv^2)^2}{4u^2v^4D}\right]\nonumber\,.
\end{align}
This is all the further we can go without an explicit form for the matrix element.

\subsubsection{$C_F$ Channel}

In the $C_F$ color channel, the matrix element for two soft gluon emission is \cite{Catani:1999ss}
\begin{equation}
|{\cal M}_{C_F}(k_1,k_2)|^2 = 16g^4 C_F^2 \frac{1}{E_1^2\sin^2\theta_1E_2^2\sin^2\theta_2}\to \frac{(4\pi\alpha_s)^2 C_F^2 }{Q^4z_1^4\theta_1^4}\frac{512}{u^2v^2}\,,
\end{equation}
where we have expanded in the collinear limit and used the changes of variables developed above.  We have also multiplied by a factor of 2 in the soft and collinear limit to account for emission off of the quark or anti-quark in the final state.  In this color channel, the matrix element is then
\begin{align}
\frac{1}{\sigma_0}\frac{d^2\sigma}{dD\, dy} &= \frac{\alpha_s^2C_F^2}{\pi^3}\frac{1}{y D}\int_0^\infty \frac{du}{u} \int_0^\infty \frac{dv}{v}\, \Theta\left(\frac{1}{y}-\frac{(1+uv^2)^2}{3uv^2}\right)\left(
\frac{1}{y}\frac{3uv^2}{(1+uv^2)^2}-1
\right)^{-1/2}\\
&
\hspace{2cm}\times\left[
\Theta(1-v)\Theta(1-u)\Theta\left(3(1+uv^2)-2\sqrt{\frac{D}{y}}\right)\log\frac{9y(1+uv^2)^2}{4D}\right.\nonumber\\
&
\hspace{2cm}
\left.
+\Theta(1-v)\Theta(u-1)\Theta\left(3(1+uv^2)-2u\sqrt{\frac{D}{y}}\right)\log\frac{9y(1+uv^2)^2}{4u^2D}\right.\nonumber\\
&
\hspace{2cm}
\left.
+\Theta(v-1)\Theta(1-u)\Theta\left(3(1+uv^2)-2v^2\sqrt{\frac{D}{y}}\right)\log\frac{9y(1+uv^2)^2}{4v^4D}\right.\nonumber\\
&
\hspace{2cm}
\left.
+\Theta(v-1)\Theta(u-1)\Theta\left(3(1+uv^2)-2uv^2\sqrt{\frac{D}{y}}\right)\log\frac{9y(1+uv^2)^2}{4u^2v^4D}\right]\nonumber\,.
\end{align}
It is then useful to make the change of variables
\begin{equation}
t=uv^2
\end{equation}
in place of the $u$ integral.  The cross section with this change of variables becomes
\begin{align}
\frac{1}{\sigma_0}\frac{d^2\sigma}{dD\, dy} &= \frac{\alpha_s^2C_F^2}{\pi^3}\frac{1}{y D}\int_0^\infty \frac{dt}{t} \int_0^\infty \frac{dv}{v}\, \Theta\left(\frac{1}{y}-\frac{(1+t)^2}{3t}\right)\left(
\frac{1}{y}\frac{3t}{(1+t)^2}-1
\right)^{-1/2}\\
&
\hspace{2cm}\times\left[
\Theta(1-v)\Theta(v^2-t)\Theta\left(3(1+t)-2\sqrt{\frac{D}{y}}\right)\log\frac{9y(1+t)^2}{4D}\right.\nonumber\\
&
\hspace{2cm}
\left.
+\Theta(1-v)\Theta(t-v^2)\Theta\left(3v^2(1+t)-2t\sqrt{\frac{D}{y}}\right)\log\frac{9yv^4(1+t)^2}{4t^2D}\right.\nonumber\\
&
\hspace{2cm}
\left.
+\Theta(v-1)\Theta(v^2-t)\Theta\left(3(1+t)-2v^2\sqrt{\frac{D}{y}}\right)\log\frac{9y(1+t)^2}{4v^4D}\right.\nonumber\\
&
\hspace{2cm}
\left.
+\Theta(v-1)\Theta(t-v^2)\Theta\left(3(1+t)-2t\sqrt{\frac{D}{y}}\right)\log\frac{9y(1+t)^2}{4t^2D}\right]\nonumber\,.
\end{align}
The integral over $v$ can be done and the integral can be massaged into a simple form, with one integral over $t\in[0,1]$ remaining:
\begin{align}\label{eq:cfcdreg3}
\frac{1}{\sigma_0}\frac{d^2\sigma}{dD\, dy} &= \frac{\alpha_s^2C_F^2}{\pi^3}\frac{1}{y D}\int_0^1 \frac{dt}{t} \, \Theta\left(\frac{1}{y}-\frac{(1+t)^2}{3t}\right)\left(
\frac{1}{y}\frac{3t}{(1+t)^2}-1
\right)^{-1/2}\\
&
\hspace{2cm}\times\Theta\left(3(1+t)-2\sqrt{\frac{D}{y}}\right)\left[
-\log t\log\frac{9y(1+t)^2}{4D}+\frac{1}{2}\log^2\frac{9y(1+t)^2}{4D}\right]\nonumber\,.
\end{align}

To isolate the regime in which $y\sim 1$ ($D\sim C^2$), we need to remove the $y\ll 1$ limit.  In this limit, $y$ and $t$ scale similarly, so we can expand the integral above to first order in $y,t\ll 1$:
\begin{align}
\frac{1}{\sigma_0}\frac{d^2\sigma^{y\ll 1}}{dD\, dy} &= \frac{\alpha_s^2C_F^2}{\pi^3}\frac{1}{y D}\int_0^\infty \frac{dt}{t} \, \Theta\left(\frac{1}{y}-\frac{1}{3t}\right)\left(
\frac{3t}{y}-1
\right)^{-1/2}\\
&
\hspace{2cm}\times\Theta\left(3-2\sqrt{\frac{D}{y}}\right)\left[
-\log t\log\frac{9y}{4D}+\frac{1}{2}\log^2\frac{9y}{4D}\right]\nonumber\\
&=\frac{\alpha_s^2C_F^2}{\pi^2}\frac{1}{y D}\left[
\log\frac{3}{4y}\log\frac{9y}{4D}+\frac{1}{2}\log^2\frac{9y}{4D}\right]\Theta\left(\frac{3}{4}-y\right)\Theta\left(y-\frac{4D}{9}\right)\nonumber\,.
\end{align}
We just need to subtract this from \Eq{eq:cfcdreg3} to determine the cross section in the region where $y\sim 1$.  It is also useful to verify that this expression agrees with the leading-logarithmic calculation of \Sec{sec:event2}.  To leading-logarithmic accuracy, we can set all numerical factors in logarithms to 1, yielding
\begin{align}
\frac{1}{\sigma_0}\frac{d^2\sigma^{y\ll 1,\text{LL}}}{dD\, dy}=\frac{\alpha_s^2C_F^2}{\pi^2}\frac{1}{y D}\left[
\log\frac{1}{y}\log\frac{y}{D}+\frac{1}{2}\log^2\frac{y}{D}\right]\Theta(1-y)\Theta(y-D)\,.
\end{align}
Then, to calculate the leading-logarithmic cross section for the $D$-parameter, we just integrate over $y$:
\begin{align}
\frac{1}{\sigma_0}\frac{d\sigma^\text{LL}}{dD} &= \frac{\alpha_s^2C_F^2}{\pi^2}\frac{1}{D}\int_D^{1}\frac{dy}{y}\,\left[
\log\frac{1}{y}\log\frac{y}{D}+\frac{1}{2}\log^2\frac{y}{D}\right]\\
&= -\frac{\alpha_s^2C_F^2}{\pi^2}\frac{\log^3 D}{3D}
\nonumber \,.
\end{align}
This agrees with the results of \Sec{sec:ll}.

We are unable to do this integral exactly, and to implement the resummation of region 3, we choose to just numerically integrate over $y$.  To compare to the ${\cal O}(\alpha_s^2)$ results from {\tt EVENT2}, we need to integrate over $y$.  To good approximation, we find the following result:
\begin{equation}
\frac{1}{\sigma_0}\frac{d^2\sigma^{\alpha_s^2,y\sim 1}}{dD} \approx -13.16\left(\frac{\alpha_s}{2\pi}\right)^2 C_F^2 \frac{\log D}{D}\,.
\end{equation}

\subsubsection{$C_A$ Channel}

In the $C_A$ color channel, the matrix element for two soft gluon emission in the soft and collinear limit is \cite{Catani:1999ss}
\begin{align}
|{\cal M}_{C_A}(k_1,k_2)|^2 &=128\frac{(4\pi\alpha_s)^2}{Q^4z_1^4\theta_1^4} C_FC_A\left[
\frac{(1-v^2)^2}{(1+v^2-2v\cos\phi)^2(1+uv^2)^2(1+u)^2}\right.\\
&
\hspace{-3cm}
\left.-\frac{1}{2u^2v^2}\left(
2-\frac{u(1+v^2)}{(1+uv^2)(1+u)}
\right)+\frac{1}{u(1+v^2-2v\cos\phi)}\left(
\frac{1}{u}+\frac{1}{uv^2}-\frac{1+6v^2+v^4}{2v^2(1+uv^2)(1+u)}\right)
\right]\nonumber \,,
\end{align}
where we have expanded in the collinear limit and used the changes of variables developed above.  We have also multiplied by 2 to account for collinear emission off of the quark or anti-quark.  As everything else in the integral is symmetric in $\cos\phi\to-\cos\phi$, we can explicitly do this symmetrization:
\begin{align}
|{\cal M}_{C_A}(\cos\phi)|^2+|{\cal M}_{C_A}(-\cos\phi)|^2 &\propto2\left[
\frac{(1-v^2)^2((1+v^2)^2+4v^2\cos^2\phi)}{((1-v^2)^2+4v^2\sin^2\phi)^2(1+uv^2)^2(1+u)^2}\right.\\
&
\hspace{-6cm}
\left.-\frac{1}{2u^2v^2}\left(
2-\frac{u(1+v^2)}{(1+uv^2)(1+u)}
\right)+\frac{1+v^2}{uv^2((1-v^2)^2+4v^2\sin^2\phi)}\left(
\frac{1+v^2}{u}-\frac{1+6v^2+v^4}{2(1+uv^2)(1+u)}\right)
\right]\,.\nonumber
\end{align}
The cross section in the $C_A$ color channel is then
\begin{align}
\hspace{-0.3cm}\frac{1}{\sigma_0}\frac{d^2\sigma}{dD\, dy} &= \frac{\alpha_s^2 C_F C_A}{4\pi^3}\frac{1}{y D}\int_0^\infty du\,u \int_0^\infty dv\, v\,  \left[
\frac{(1-v^2)^2((1+v^2)^2+4v^2\cos^2\phi)}{((1-v^2)^2+4v^2\sin^2\phi)^2(1+uv^2)^2(1+u)^2}\right.\\
&
\hspace{-1cm}
\left.-\frac{1}{2u^2v^2}\left(
2-\frac{u(1+v^2)}{(1+uv^2)(1+u)}
\right)+\frac{1+v^2}{uv^2((1-v^2)^2+4v^2\sin^2\phi)}\left(
\frac{1+v^2}{u}-\frac{1+6v^2+v^4}{2(1+uv^2)(1+u)}\right)
\right]\nonumber\\
&
\hspace{4cm}
\times \Theta\left(\frac{1}{y}-\frac{(1+uv^2)^2}{3uv^2}\right)\left(
\frac{1}{y}\frac{3uv^2}{(1+uv^2)^2}-1
\right)^{-1/2}\nonumber\\
&
\hspace{2cm}\times\left[
\Theta(1-v)\Theta(1-u)\Theta\left(3(1+uv^2)-2\sqrt{\frac{D}{y}}\right)\log\frac{9y(1+uv^2)^2}{4D}\right.\nonumber\\
&
\hspace{2cm}
\left.
+\Theta(1-v)\Theta(u-1)\Theta\left(3(1+uv^2)-2u\sqrt{\frac{D}{y}}\right)\log\frac{9y(1+uv^2)^2}{4u^2D}\right.\nonumber\\
&
\hspace{2cm}
\left.
+\Theta(v-1)\Theta(1-u)\Theta\left(3(1+uv^2)-2v^2\sqrt{\frac{D}{y}}\right)\log\frac{9y(1+uv^2)^2}{4v^4D}\right.\nonumber\\
&
\hspace{2cm}
\left.
+\Theta(v-1)\Theta(u-1)\Theta\left(3(1+uv^2)-2uv^2\sqrt{\frac{D}{y}}\right)\log\frac{9y(1+uv^2)^2}{4u^2v^4D}\right]\nonumber\,.
\end{align}
We need to make the replacement then that
\begin{equation}
\sin^2\phi=y\frac{(1+uv^2)^2}{3uv^2}\,,
\end{equation}
as well as the change of variables
\begin{equation}
t=uv^2\,.
\end{equation}
The cross section is then
\begin{align}
\hspace{-0.5cm}\frac{1}{\sigma_0}\frac{d^2\sigma}{dD\, dy} &= \frac{\alpha_s^2 C_F C_A}{4\pi^3}\frac{1}{y D}\int_0^\infty dt\, t \int_0^\infty \frac{dv}{v^3}\,  \left[
v^4\frac{(1-v^2)^2\left((1+v^2)^2+4v^2\left(1-y\frac{(1+t)^2}{3t}\right)\right)}{\left((1-v^2)^2+4yv^2\frac{(1+t)^2}{3t}\right)^2(1+t)^2(v^2+t)^2}\right.\\
&
\hspace{-1cm}
\left.-\frac{v^2}{2t^2}\left(
2-\frac{t(1+v^2)}{(1+t)(v^2+t)}
\right)+\frac{v^2(1+v^2)}{t\left((1-v^2)^2+4yv^2\frac{(1+t)^2}{3t}\right)}\left(
\frac{1+v^2}{t}-\frac{1+6v^2+v^4}{2(1+t)(v^2+t)}\right)
\right]\nonumber\\
&
\hspace{4cm}
\times \Theta\left(\frac{1}{y}-\frac{(1+t)^2}{3t}\right)\left(
\frac{1}{y}\frac{3t}{(1+t)^2}-1
\right)^{-1/2}\nonumber\\
&
\hspace{2cm}\times\left[
\Theta(1-v)\Theta(v^2-t)\Theta\left(3(1+t)-2\sqrt{\frac{D}{y}}\right)\log\frac{9y(1+t)^2}{4D}\right.\nonumber\\
&
\hspace{2cm}
\left.
+\Theta(1-v)\Theta(t-v^2)\Theta\left(3v^2(1+t)-2t\sqrt{\frac{D}{y}}\right)\log\frac{9yv^4(1+t)^2}{4t^2D}\right.\nonumber\\
&
\hspace{2cm}
\left.
+\Theta(v-1)\Theta(v^2-t)\Theta\left(3(1+t)-2v^2\sqrt{\frac{D}{y}}\right)\log\frac{9y(1+t)^2}{4v^4D}\right.\nonumber\\
&
\hspace{2cm}
\left.
+\Theta(v-1)\Theta(t-v^2)\Theta\left(3(1+t)-2t\sqrt{\frac{D}{y}}\right)\log\frac{9y(1+t)^2}{4t^2D}\right]\nonumber\,.
\end{align}

To restrict to the region where $y\sim 1$, we need to subtract the limit in which $y\ll1$.  Unlike the $C_F$ color channel, this expansion in the $C_A$ channel is challenging, so we just extract the limiting behavior numerically.  We find
\begin{align}
\hspace{-0.3cm}
\frac{1}{\sigma_0}\frac{d^2\sigma^{y\ll 1}}{dy\, dD} &=\frac{\alpha_s^2C_FC_A}{2\pi^2}\frac{1}{yD}\left[
\log\frac{9y}{4D}\log \frac{3}{4y}- \log \frac{9y}{4D}-\frac{3}{2}
\right]\Theta\left(
\frac{3}{4}-y
\right)\Theta\left(
y-\frac{4}{9}D
\right)\,.
\end{align}
The leading logarithmic term agrees exactly with the expression from the region 1/2 factorization theorem derived in \Sec{sec:ll}.  Subtracting this from the full expression then restricts to $y\sim 1$.  We are unable to do this integral exactly, and to implement the resummation of region 3, we choose to just numerically integrate over $y$.  We find the following approximate result:
\begin{equation}
\frac{1}{\sigma_0}\frac{d^2\sigma^{\alpha_s^2,y\sim 1}}{dD} \approx -1.4\left(\frac{\alpha_s}{2\pi}\right)^2 C_FC_A \frac{\log D}{D}\,.
\end{equation}

\subsubsection{$n_f$ Channel}

In the $n_f$ color channel, the matrix element for two soft quark emission in the soft and collinear limit is \cite{Catani:1999ss}
\begin{equation}
|{\cal M}_{n_f}(k_1,k_2)|^2 =256\frac{(4\pi\alpha_s)^2}{Q^4z_1^4\theta_1^4} C_Fn_fT_R\frac{(1+uv^2)^2+v^2(1+u)^2-2v(1+u)(1+uv^2)\cos\phi}{u(1+v^2-2v\cos\phi)^2(1+uv^2)^2(1+u)^2} \,,
\end{equation}
where we have expanded in the collinear limit and used the changes of variables developed above.  We have also multiplied by 2 to account for collinear emission off of the quark or anti-quark.  As everything else in the integral is symmetric in $\cos\phi\to-\cos\phi$, we can explicitly do this symmetrization:
\begin{align}
&|{\cal M}_{n_f}(\cos\phi)|^2+|{\cal M}_{n_f}(-\cos\phi)|^2\\
&
\hspace{0cm}
\propto 2\frac{(1-v^2)^2\left(1+v^2(1-4u+u^2)+u^2v^4\right)+4v^2\left(2u(1+v^4)+(1+u^2 v^2)(1+v^2)^2\right)\sin^2\phi}{u((1-v^2)^2+4v^2\sin^2\phi)^2(1+uv^2)^2(1+u)^2} \,.\nonumber
\end{align}
In this color channel, the cross section is then
\begin{align}
\frac{1}{\sigma_0}\frac{d^2\sigma}{dD\, dy} &= \frac{\alpha_s^2C_F n_f T_R}{2\pi^3}\frac{1}{y D}\int_0^\infty du \int_0^\infty dv\, \Theta\left(\frac{1}{y}-\frac{(1+uv^2)^2}{3uv^2}\right)\left(
\frac{1}{y}\frac{3uv^2}{(1+uv^2)^2}-1
\right)^{-1/2} \\
&
\hspace{-1cm}
\times v\frac{(1-v^2)^2\left(1+v^2(1-4u+u^2)+u^2v^4\right)+4v^2\left(2u(1+v^4)+(1+u^2 v^2)(1+v^2)^2\right)\sin^2\phi}{((1-v^2)^2+4v^2\sin^2\phi)^2(1+uv^2)^2(1+u)^2}\nonumber\\
&
\hspace{2cm}\times\left[
\Theta(1-v)\Theta(1-u)\Theta\left(3(1+uv^2)-2\sqrt{\frac{D}{y}}\right)\log\frac{9y(1+uv^2)^2}{4D}\right.\nonumber\\
&
\hspace{2cm}
\left.
+\Theta(1-v)\Theta(u-1)\Theta\left(3(1+uv^2)-2u\sqrt{\frac{D}{y}}\right)\log\frac{9y(1+uv^2)^2}{4u^2D}\right.\nonumber\\
&
\hspace{2cm}
\left.
+\Theta(v-1)\Theta(1-u)\Theta\left(3(1+uv^2)-2v^2\sqrt{\frac{D}{y}}\right)\log\frac{9y(1+uv^2)^2}{4v^4D}\right.\nonumber\\
&
\hspace{2cm}
\left.
+\Theta(v-1)\Theta(u-1)\Theta\left(3(1+uv^2)-2uv^2\sqrt{\frac{D}{y}}\right)\log\frac{9y(1+uv^2)^2}{4u^2v^4D}\right]\nonumber\,.
\end{align}

We need to make the replacement then that
\begin{equation}
\sin^2\phi=y\frac{(1+uv^2)^2}{3uv^2}\,,
\end{equation}
as well as the change of variables
\begin{equation}
t=uv^2\,.
\end{equation}
The cross section is then
\begin{align}
\frac{1}{\sigma_0}\frac{d^2\sigma}{dD\, dy} &= \frac{\alpha_s^2C_F n_f T_R}{2\pi^3}\frac{1}{y D}\int_0^\infty dt \int_0^\infty dv\, \Theta\left(\frac{1}{y}-\frac{(1+t)^2}{3t}\right)\left(
\frac{1}{y}\frac{3t}{(1+t)^2}-1
\right)^{-1/2} \\
&
\hspace{-1cm}
\times \frac{(1-v^2)^2\left(1+v^2(1-4\frac{t}{v^2}+\frac{t^2}{v^4})+t^2\right)+4v^2\left(2\frac{t}{v^2}(1+v^4)+(1+\frac{t^2}{v^2})(1+v^2)^2\right)y\frac{(1+t)^2}{3t}}{v((1-v^2)^2+4v^2y\frac{(1+t)^2}{3t})^2(1+t)^2(1+\frac{t}{v^2})^2}\nonumber\\
&
\hspace{2cm}\times\left[
\Theta(1-v)\Theta(v^2-t)\Theta\left(3(1+t)-2\sqrt{\frac{D}{y}}\right)\log\frac{9y(1+t)^2}{4D}\right.\nonumber\\
&
\hspace{2cm}
\left.
+\Theta(1-v)\Theta(t-v^2)\Theta\left(3v^2(1+t)-2t\sqrt{\frac{D}{y}}\right)\log\frac{9yv^4(1+t)^2}{4t^2D}\right.\nonumber\\
&
\hspace{2cm}
\left.
+\Theta(v-1)\Theta(v^2-t)\Theta\left(3(1+t)-2v^2\sqrt{\frac{D}{y}}\right)\log\frac{9y(1+t)^2}{4v^4D}\right.\nonumber\\
&
\hspace{2cm}
\left.
+\Theta(v-1)\Theta(t-v^2)\Theta\left(3(1+t)-2t\sqrt{\frac{D}{y}}\right)\log\frac{9y(1+t)^2}{4t^2D}\right]\nonumber\,.
\end{align}

To restrict to the region where $y\sim 1$, we need to subtract the limit in which $y\ll1$.  Unlike the $C_F$ color channel, this expansion in the $n_f$ channel is challenging, so we just extract the limiting behavior numerically.  We find
\begin{align}
\hspace{-0.3cm}
\frac{1}{\sigma_0}\frac{d^2\sigma^{y\ll 1}}{dy\, dD} &=\frac{\alpha_s^2C_Fn_f T_R}{\pi^2}\frac{\log\frac{9y}{4D}}{6yD}
\Theta\left(
\frac{3}{4}-y
\right)\Theta\left(
y-\frac{4}{9}D
\right)\nonumber\,.
\end{align}
When integrated over $y$, the leading logarithmic term agrees exactly with the expression from the region 1$-$2 factorization theorem derived in \Sec{sec:ll}.  Subtracting this from the full expression then restricts to $y\sim 1$.  We are unable to do this integral exactly, and to implement the resummation of region 3, we choose to just numerically integrate over $y$.  We find the following approximation:
\begin{equation}
\frac{1}{\sigma_0}\frac{d^2\sigma^{\alpha_s^2,y\sim 1}}{dD} \approx -0.31 \left(\frac{\alpha_s}{2\pi}\right)^2 C_Fn_f T_R \frac{\log D}{D}\,.
\end{equation}

\bibliography{d_param}

\end{document}